\documentclass[10pt,article]{IEEEtran}
\usepackage{graphicx}
\usepackage{multirow}
\usepackage{amsmath}
\usepackage{times}
\newcommand{\mr}[1]{\multirow{2}*{#1}}

\newcommand{\RelaxFloats}{
       \renewcommand{\topfraction}{0.9}
       \renewcommand{\floatpagefraction}{0.9}
       \renewcommand{\textfraction}{0.1}
}

\begin{document}

\RelaxFloats


\title{Mining Behavioral Groups in Large Wireless LANs}

\author{\authorblockN{Wei-jen Hsu$^{1}$, Debojyoti Dutta$^{2}$\thanks{This work was done when Debojyoti Dutta was with University of Southern California.}, and Ahmed Helmy$^{1}$}
\authorblockA{\\$^{1}$Department of Computer and Information Science and Engineering, University of Florida $^{2}$Cisco Systems, Inc.
\\Email: $^{1}$ \{wjhsu, helmy\}@ufl.edu, $^{2}$dedutta@cisco.com}}

\maketitle

\begin{abstract}

One vision of future wireless networks is that they will be deeply integrated and embedded in our lives and will involve the use of personalized mobile devices. User behavior in such networks is bound to affect the network performance. It is imperative to study and characterize the fundamental structure of wireless user behavior in order to model, manage, leverage and design efficient mobile networks. It is also important to make such study as realistic as possible, based on extensive measurements collected from existing deployed wireless networks.

In this study, using our systematic {\it TRACE} approach, we analyze wireless users' behavioral patterns by extensively mining wireless network logs from two major university campuses. We represent the data using location preference vectors, and utilize unsupervised learning (clustering) to classify trends in user behavior using novel similarity metrics. Matrix decomposition techniques are used to identify (and differentiate between) major patterns. While our findings validate intuitive repetitive behavioral trends and user grouping, it is surprising to find the qualitative commonalities of user behaviors from the two universities. We discover multi-modal user behavior for more than $60\%$ of the users, and there are hundreds of distinct groups with unique behavioral patterns in both campuses. The sizes of the major groups follow a power-law distribution. Our methods and findings provide an essential step towards network management and behavior-aware network protocols and applications, to name a few.


\end{abstract}

\section{Introduction} \label{intro}

In recent years, we have witnessed the mass deployments of portable computing and communication devices (e.g., cellphones, laptops, PDAs) and wireless communication infrastructures. As the adoption of these technologies becomes an inseparable part of our lives, we envision that future usage of mobile
devices and services will be highly personalized. Users will
incorporate these new technologies into their daily lives, and the way
they use new devices and services will reflect their personality and
lifestyle truthfully. To understand its impact, we believe
that there is a pressing need to go beyond the technological
perspective and capture and understand the user behavioral patterns as
users adopt the new technology. This understanding will also play a crucial role in solving a multitude of technical issues, ranging from better network management to designing of behavior-aware protocols, services, and user models.

Consider wireless LANs (WLANs) on  university campuses as an example. One could imagine the major work places (e.g., offices and classrooms) and the informational hubs  (e.g., libraries and computer centers) would dominate users' behavioral patterns in terms of network usage. However, as the WLAN deployments prevail, the location from  where people access information is bound to change. While the traditional "hot spots" still play an important role, we can expect users to display diverse behavioral patterns that reflect their personal preferences (e.g., A small group may prefer to work at a coffee shop), as these wireless devices become tiny and personalized. We need to understand such behavioral patterns to better characterize the users within a social context. The technique to discover such patterns from collected data is the focus of our paper.

In this paper we take a first step towards understanding and characterizing the structure of behavioral patterns of users within large WLANs. We develop methods to identify groups of users that demonstrate similar and coherent behavioral pattern. This is important for several reasons: (1) From the network management perspective, it helps us to  understand the potential interplay of the user groups with the network operation and reveals insight previously unavailable by looking at the mere aggregate network statistics. (2) From the application or service perspective, the groups identify different existing major behavioral modes in the network, and, hence, can be potentially utilized to identify targets for group-aware services. (3) From a social sciences perspective, the results unravel the relationships between users (i.e., their "closeness" in terms of network usage behavior) when they incorporate wireless mobile devices as an inseparable part of their daily lives.

We apply our analysis framework on long-term WLAN traces obtained from two university campuses\cite{MobiLib-data, Dart-movement-data} across the coasts of USA. We represent a user's behavioral features by constructing the normalized association matrix to which we apply our analysis. While the applicability of our methods is not specific to WLANs, these are the most extensive wireless user behavioral traces available today. Although this is not the first study of these WLAN traces, our unique focus is on user groups across campuses while most of the previous studies focus on individual user behavior models or aggregate statistics within one campus. We leverage unsupervised learning (i.e., clustering) techniques \cite{cluster-tech} to determine groups of users displaying similar behavior. While clustering has been widely-applied in other areas, the main contribution of the paper is to construct proper representations for our data sets and design novel distance metrics between users. These two aspects are fundamental in the application of clustering techniques and determine the quality of the results we obtain. The key challenge in designing a good distance metric is to accurately and succinctly summarize the trends in the data, so the distances are not influenced by noise and can be evaluated efficiently. We show that a singular-value decomposition (SVD) based scheme not only provides the best summary of the data, but also leads to a distance metric that is robust to noise and computationally efficient. Furthermore, we validate our methods and explain its significance.

We find the following common trends from the two diverse datasets: (1) More than $60\%$ of the WLAN users display multi-modal behavior (their behavior can be decomposed into multiple modes or types) in the long run. However, for many users the most dominant behavioral mode is much stronger than the rest. This leads to efficient summaries of their behavioral patterns. With SVD, we can capture more than $90\%$ of the power in the association patterns with just five components. (2) Current university WLANs consist of a large number of user groups with distinct association patterns, in the order of hundreds. We find that the distributions of sizes of the major groups, however, are highly skewed and follow a power-law distribution. The top-$10$ groups contain at least $33\%$ of the users while about a half of the identified groups have less than $10$ members. It is surprising to find qualitative commonalities in user behavior almost across the board considering the differences (e.g., Geographical locations, sizes and structures of the campuses, different student bodies, etc.) among the campuses.

\begin{figure}

\centering

\includegraphics[width=2.8in]{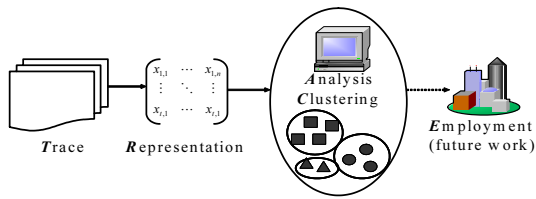}

\caption{Illustration of the {\it TRACE} approach.}
\label{approach}
\end{figure}

We use Fig. \ref{approach} to illustrate the conceptual flow of our approach in the paper, which we refer to as the {\it TRACE} approach. The five major components in the approach are: {\it T}race, {\it R}epresentation, {\it A}nalysis, {\it C}lustering, and {\it E}mployment (or application). The work starts with the WLAN {\it traces} that capture realistic user behavior. We then focus on a specific {\it representation} distilled from the traces that captures important aspects of user behavior, as we introduce in section \ref{prelim}. We conduct {\it analysis} and {\it clustering} upon the users based on the chosen representation, {\it normalized association vectors}, from section \ref{naive} to section \ref{interpretation}. We first show the need for a good distance metric for clustering in section \ref{naive}. To achieve that goal, we conduct further {\it analysis} to understand the nature of user association patterns, and evaluate and contrast various summaries to capture its major trend in section \ref{summary-methods}. We then utilize a feature-based approach to achieve meaningful user {\it clustering} in section \ref{feature} and discuss its interpretation in section \ref{interpretation}. Finally, we show-case one direct application, {\it mobility-profile-based casting}, of our user grouping in section \ref{MPcast}, and briefly discuss other potential {\it employment} of the methods and findings in the paper in section \ref{disc}. Related work is discussed in section \ref{rw}. The paper concludes in section \ref{conclusion}.

\section{Preliminaries} \label{prelim}

We first introduce the traces we analyze in the paper and the {\it
normalized association vector} representation we choose. We also
briefly introduce the necessary background knowledge about clustering
in the section.

\subsection{Choice of Data Set and Representations} \label{representation}


The widespread deployments of large-scale wireless LANs on university
campuses have attracted high adoption from its community. These 
deployments have outgrown experimental networks and
become a commodities. 
Due to its high penetration and diversity in users (as compared to
corporate WLANs), campus networks are good platforms to study
the behavioral pattern of WLAN users. To our benefit, great efforts
have already been made to collect the user traces from several large
WLAN deployments \cite{MobiLib-web, CRAWDAD-web}. We elect two WLAN
traces collected from large populations for long durations for the
study. The details for the selected traces are listed in Table
\ref{trace-facts}.

\begin{table}
\caption{Facts about studied traces}
\label{trace-facts}
\begin{center}
\begin{tabular}{|c||c|c|}
\hline
Trace source & USC \cite{MobiLib-data} & Dartmouth \cite{Dart-movement-data} \\
\hline
Time/duration & 2006 spring & 2004 spring  \\
 of trace & semester & quarter \\
 & (94 days) & (61 days) \\
\hline
Start/End & 01/25/06- & 04/05/04- \\
time & 04/28/06 & 06/04/04 \\
\hline
Location & \mr{Building} & \mr{Access point} \\
granularity & &\\
\hline
Unique & \mr{137 buildings} & 545 APs/ \\
locations & & 162 buildings \\
\hline 
Unique MACs analyzed & 5,000 & 6,582 \\
\hline
\end{tabular}

\end{center}
\end{table}

While university WLAN traces are suitable for the study of user
behavior, there are also shortcomings in these traces. The most
important ones are (1) Users are not always online and many of them
access the network sporadically. (2) Most WLAN users access the
network with laptops, which are not always easily portable and limit the
mobility of users while accessing the network. 
However, these WLAN
traces are by far the most extensive publicly available traces and we can 
indeed discover interesting patterns. 
Note that our methods are not limited to the specific data sets we
choose, and it would be of great interest to study traces from other
mobile devices (e.g., cellphones, iPods), if available for a large
population.


To understand user behavior from wireless network traces, the first
fundamental task is to choose a representation of the raw data. 
We choose the patterns of users visiting various locations in the WLAN 
for the analysis.
Visiting pattern is important to WLANs as mobility is one of
its defining characteristics. When a WLAN user moves within the campus
and {\it associates} with access points (APs) across the network, the
set of APs with which the user associates is considered an indicator
of the user's physical location. From a social context, the places a
person visits regularly and repeatedly usually have a stronger
connection to her identity and affiliation. It is perhaps one of the
important distinguishing factors for people with different social
attributes. 

\begin{figure}

\centering

\includegraphics[width=2.8in]{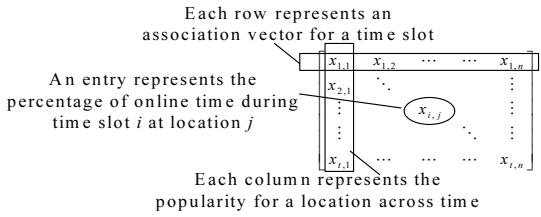}

\caption{Illustration of association matrix representation.}
\label{matrix_illustration}
\end{figure}

We represent a user's visiting pattern by what we refer to as {\it
normalized association vectors}\footnote{For brevity, we sometimes use the
shortened term {\it association vector} to refer to {\it normalized association
vector} unless stated otherwise.}.
The association vector is a summary of a user's 
association with various locations during a given time slot.  We choose to use a day as the
time slot since it represents the most natural behavior cycle in our
lives. The {\it association vector} for each time slot is an $n$-entry
vector, $(x_1, x_2, ..., x_n)$, where $n$ is the number of unique locations
(i.e., buildings) in the given trace.
Note that, although WLAN traces provide better location granularity 
(at per-access point level), in this work we aggregate APs in the same building as
a single location for better interpretation of user behavior.
Each entry in the vector, $x_i$, represents the {\it
fraction} of online time the user spends at the location during the
time slot, i.e. we normalize the user association time with respect to
his online time. With this representation, the conclusions we draw are not influenced by
the absolute value of online time, which varies across a wide range
among different users and different time slots of a given user.  Note
that the sum of the entries in the association vector, $\sum_{i=1}^{n}
x_i$, is always $1$ if the user has been online during the time
slot. We use a zero vector to represent the association vector when
the user is completely offline for the time slot.  To represent a
user's association preference for the long run, we construct the {\it
association matrix} $X$ for the user, as illustrated in
Fig. \ref{matrix_illustration}, i.e. we concatenate the association
vectors for each time slot (day).  If there are $n$ distinct locations and
the trace period consists $t$ time slots, the {\it association matrix}
for a user is a $t$-by-$n$ matrix.

Note that there are potentially many ways to represent user behavior
from a rich data set. Different representations certainly provide
different insights. Due to space limitations, we focus on the {\it normalized} representation
for {\it daily association vectors} to illustrate our analysis, and briefly
discuss about other alternatives in section \ref{alternatives}.

\subsection{Preliminaries of Clustering Techniques}

Clustering (one of the key methods in unsupervised learning) is a
widely-applied technique to discover patterns from data sets with
unknown characteristics.  It can be roughly classified into
hierarchical or partitional schemes \cite{cluster-tech}. In this paper
we use the hierarchical clustering, in which each element is initially
considered as a cluster containing one member. Then, at each step,
based on the distances between the clusters\footnote{Among several
alternatives, we use the average distance of all
element pairs between the clusters. Use of other methods
does not change the results significantly.}, two
clusters that are the closest to each other among all cluster pairs
are merged into one cluster with larger membership. This process continues until a {\it
clustering threshold} has been reached, when all the inter-cluster
distances for the remaining clusters are larger than a given distance
threshold, or the remaining cluster number reaches a given target.


One major issue in applying clustering to a data set with unknown
characteristics is that it is hard to pre-select a proper clustering
threshold in advance. The indication of a good clustering result is
that the distances between elements in the same cluster are low, and
the distances between elements in different clusters are high. (i.e.,
there is a clear separation between inter-cluster and intra-cluster
distance distributions.) Usually the clustering threshold comes from
the domain knowledge or trial-and-error. Often the decisive factor for the
quality of the clustering results is the selection of the {\it distance
metric}, which is our main contributions.

\section{Challenges}  \label{naive}


As mentioned previously, the most important step in clustering is to
define the {\it similarity} or {\it distance metric} between
users\footnote{$d(x,y)$ is a distance function if $d(x,x)=0$ and $d(x,y)$ is small if $x$ and $y$ are similar and large otherwise.
Similarity can be considered to be the opposite of distance i.e. $sim(x,y)=0$ means $x$,$y$ are dissimilar.}. We highlight the
challenges in selecting a proper distance metric with an example in
this section.

An intuitive distance function  between user association
patterns of two individuals is to consider all the association vector pairs.
Formally, we define the {\it average minimum vector distance (AMVD)} between users
$A$ and $B$, $AMVD(A,B)$, as
\begin{equation}
AMVD(A,B) = \frac{1}{\vert A \vert} \sum_{\forall A_i \in A} {\mathbf \arg \:\; \min_{\forall B_j \in B}} d(A_i,B_j),
\label{set-dist}
\end{equation}
\noindent where $A_i$ and $B_j$ denote an association vector of user
$A$ and $B$, respectively. $\vert A \vert$ denotes the cardinality of set $A$.
$d(A_i, B_j)$ denotes the Manhattan distance, defined as\footnote{
We use Manhattan distance, or the $L1$ norm,
since it is robust to statistical noise.
Note that by our representation, $0 \leq d(a,b) \leq 2$ for normalized association
vectors $a$ and $b$.}
\begin{equation} d(a,b) = \sum_{i=1}^{n} \vert a_i-b_i \vert, 
\label{man-distance}
\end{equation}
\noindent where $a_i$ and $b_i$ are the $i$-th element in vector $a$ and $b$,
respectively.
$AMVD(A,B)$ is the average of distances
from each of the vectors in set $A$ to the closest vector (or the nearest neighbor) in set
$B$. We define the $AMVD$ following the intuition that if every association
vector in set $A$ is close to some association vector in set $B$, these sets
should be similar. Note that, with this definition,
$AMVD(A,B)$ is not necessarily equal to $AMVD(B,A)$. We define a
symmetric distance metric between users $A$ and
$B$ as $D(A,B)= (AMVD(A,B)+AMVD(B,A))/2$.

\begin{figure}
\centering

\includegraphics[width=2.2in]{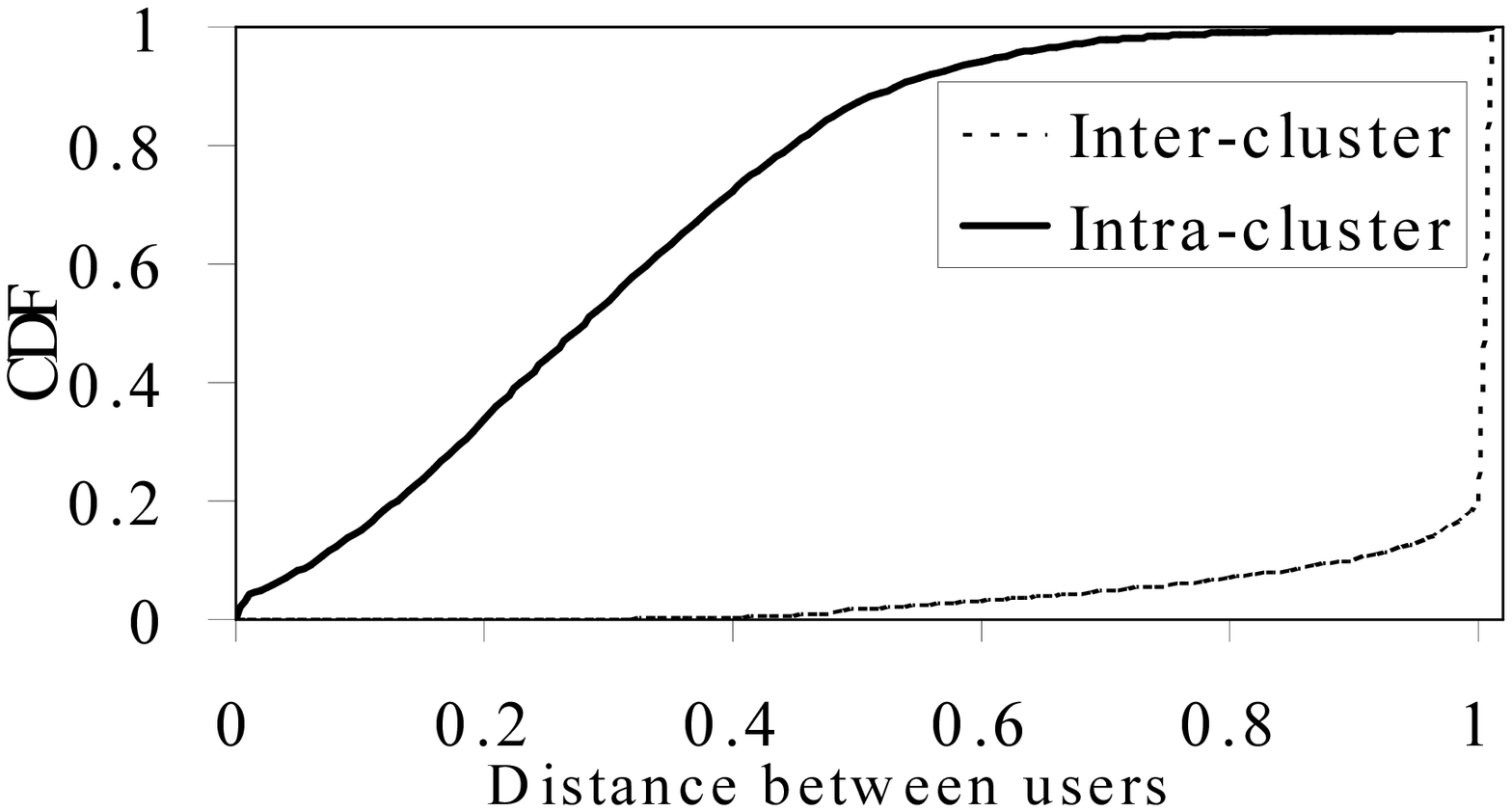} 

\footnotesize{(a) USC.}

\includegraphics[width=2.2in]{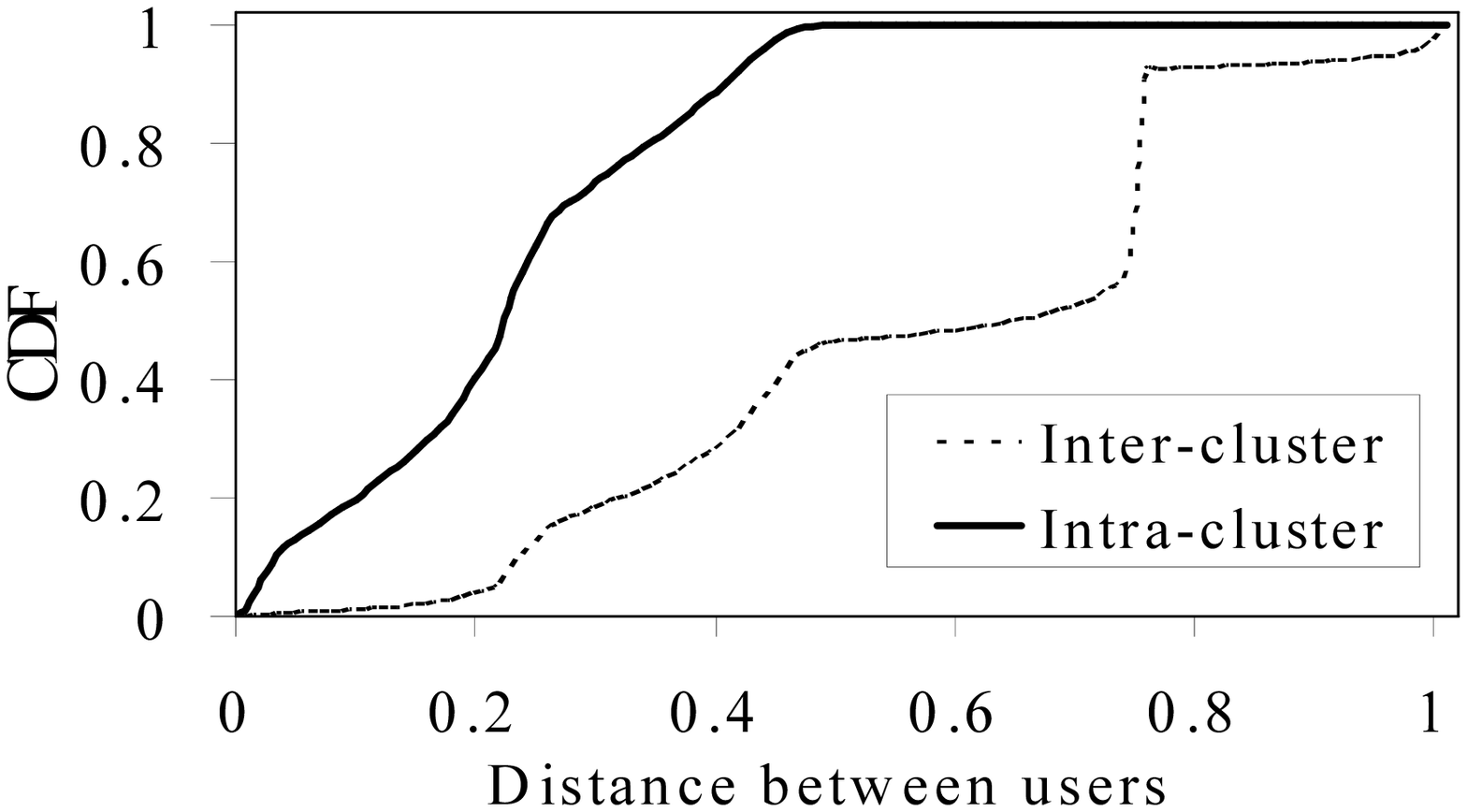} 

\footnotesize{(b) Dartmouth}

\caption{Cumulative distribution function of distances for inter-cluster and intra-cluster user pairs (AMVD distance).}

\label{inter-intra-dist-avgmind}
\end{figure}

We apply the hierarchical clustering algorithm to users with the distance metric derived from AMVD. As mentioned earlier, a clustering algorithm requires properly chosen thresholds, and the particular
choice is data-dependent.  We experiment with various thresholds, and
discover that for the USC trace, we can group the populations into
$200$ clusters with a clear separation between inter and intra cluster
distance distributions (Fig. \ref{inter-intra-dist-avgmind}
(a)), which is a qualitative indicator for a right clustering. However, the distance metric works poorly for the Dartmouth
trace, as shown in Fig. \ref{inter-intra-dist-avgmind} (b).  The
separation between inter and intra cluster distance distributions
is not clear, {\it regardless} of cluster thresholds.

One problem with the $AMVD$ metric is that it considers all association
vectors, i.e. it includes not only the important trends,
but also the noise vectors when the users deviate from the dominant
trend, leading to bad clustering results. A meaningful distance metric
should capture the major trends of user behavior and be robust to noise and outliers. 
Another problem the $AMVD$ metric is its computation complexity.
We have to calculate the distances between all $t^2$ pairs 
of association vectors for each user pair. If there are $N$ users
the computation requirement is of order $O(N^2t^2)$. Furthermore,
it requires significant space to store $t$ association vectors for all $N$ users.
Thus we would like to design a metric that is both (1) robust to noise and
(2) computation and storage efficient. In order to
achieve both goals, we start by studying the
characteristics of the association patterns of a single user to validate 
the repetitive patterns or modes of behavior. We show that this study leads us to the appropriate distance metric.

\section{Summarizing the Association Patterns} \label{summary-methods}

In this section, we identify association trends of an individual and construct a
compact representation of her association matrices, which is suitable for distance
computations used in clustering.


\subsection{Characteristics of Association Patterns}  \label{user-behavior}

We first understand the repetitive trend in a single user's
associations pattern, and how dominant the trend is (i.e., are there dominant {\it behavioral modes}?). We obtain this upon clustering the association vectors of a single individual.


Consider the clustering of the association vectors, $X_i$ for $i=1,
..., t$ (i.e., row vectors of an association matrix $X$) of a single
user. The identified clusters represent distinct {\it behavioral modes} of the user. Similar association vectors will be merged into a
cluster in the process and the cluster size
indicates its dominance - large clusters imply that the user follows consistent
association patterns on many different days as its major behavioral modes.

\begin{figure}

\centering
\includegraphics[width=2.5in]{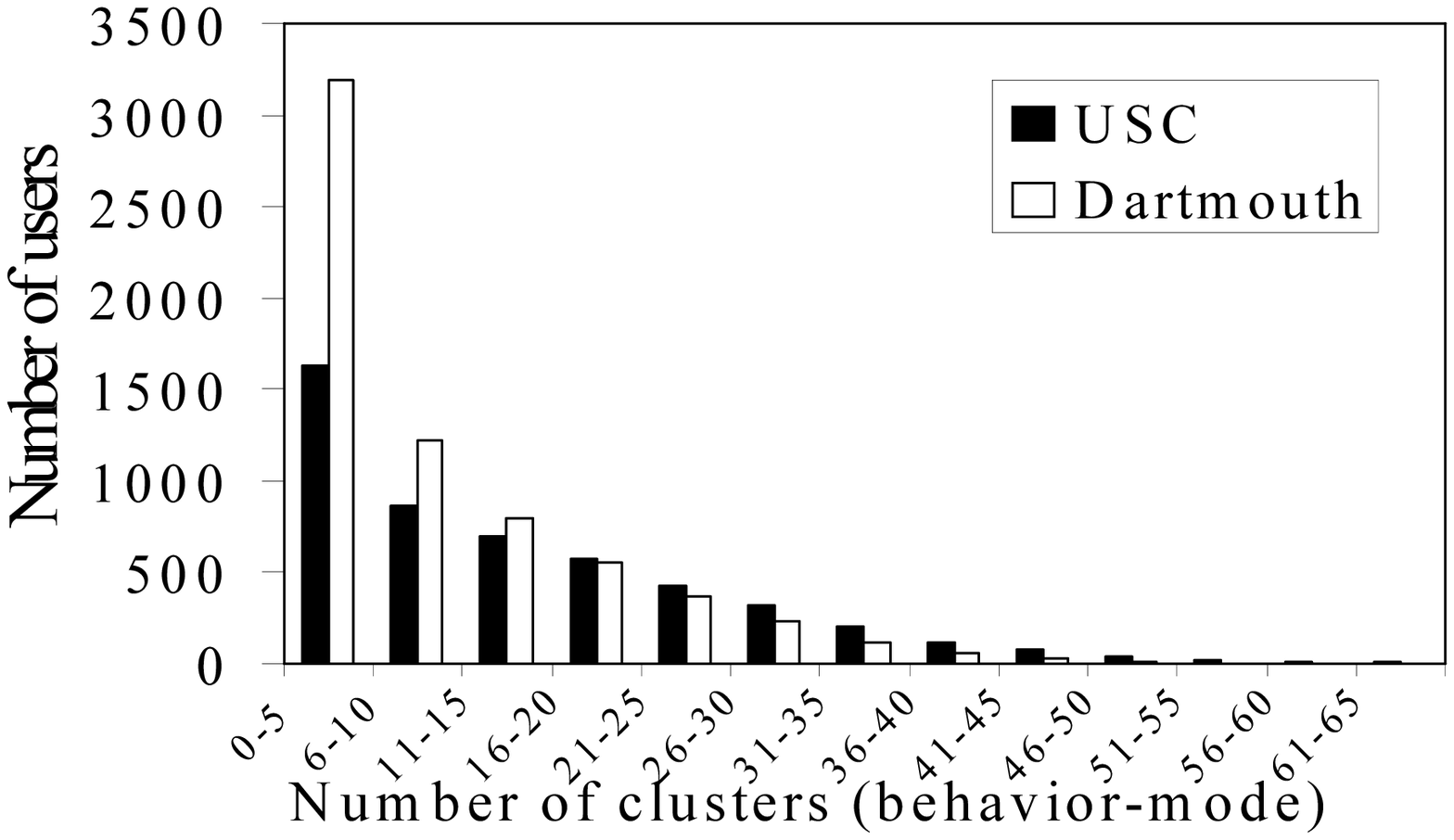} 

\footnotesize{(a) Clustering threshold = 0.2}

\includegraphics[width=2.5in]{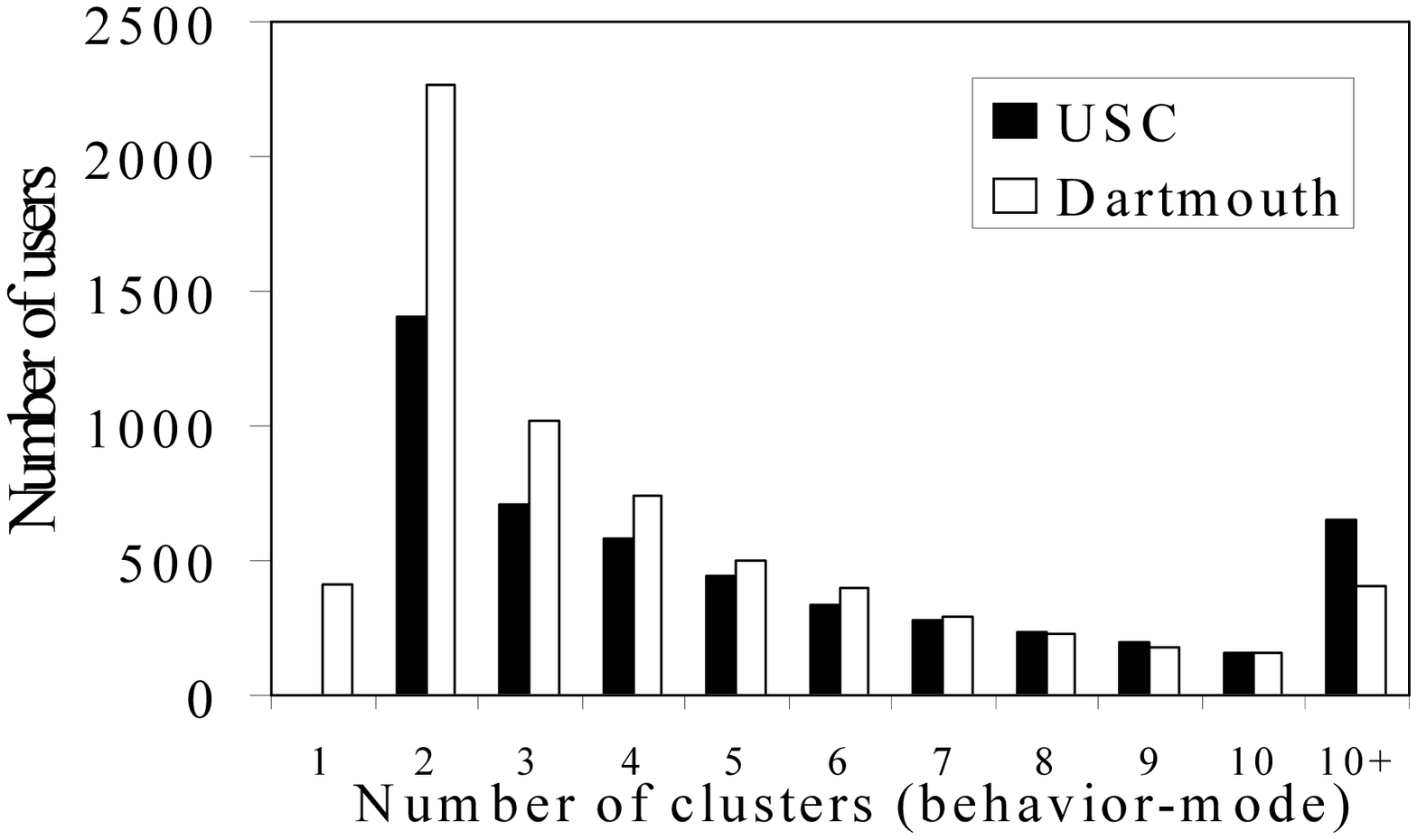} 

\footnotesize{(b) Clustering threshold = 0.9}

\caption{Distribution of number of clusters (behavioral modes) for users.}
\label{behavior_mode}
\end{figure}

We apply clustering to the association vectors of each user in the USC and the Dartmouth
traces using various clustering thresholds. The distribution of number
of clusters (or {\it behavioral modes}) obtained are
shown in Fig. \ref{behavior_mode}. In Fig. \ref{behavior_mode}(a), we
use a small clustering threshold ($0.2$), with which only very similar
association vectors are merged. We see that for the USC and the
Dartmouth traces, respectively, about $50\%$ and $67\%$ of users have
less than $10$ different clusters or behavioral modes (much fewer than
total number of time slots, $94$ and $61$) with this low clustering
threshold. This indicates the users have distinct repetitive trends
in its association vectors. 
On the other hand, if we
consider a moderate clustering threshold ($0.9$), we see in
Fig. \ref{behavior_mode} (b) that users still show multiple behavioral
modes.  On average, with $0.9$ as the clustering threshold, the number of
behavioral modes for USC and Dartmouth users are $5.57$ and $4.32$,
respectively, and the users with the most behavioral modes have $32$
clusters in both cases.

Most of those users with two behavioral modes have a consistent
association pattern: One mode corresponds to the association vectors when the user
is offline, and the other one corresponds to the association vectors when the user
is online. These users switch between online and offline behaviors
from day to day, and when they are online, the association vectors are
consistent and fall in a single behavioral mode. We refer to these
users as {\it single-modal} users.  On the other hand, we also observe
many {\it multi-modal users}. These users show a more complex
behavior: their association vectors form more than two clusters, which
indicate that they display distinct behavioral modes when they are
online. $71.9\%$ of users in USC and $59.4\%$ of users in
Dartmouth are classified as multi-modal when the clustering threshold is $0.9$. Hence, we conclude that although users in WLANs are not extremely
mobile, they do move and display various association patterns over a period of time.

\begin{figure}

\centering
\includegraphics[width=2.5in]{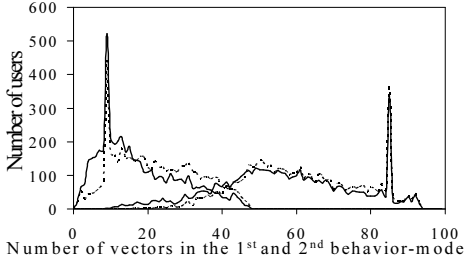} 

\caption{Distribution of association vectors in the first and the second behavioral modes for the USC trace. Right: the first cluster, Left: the second cluster.}
\label{cluster_1vs2}
\end{figure}

To examine the degree of dominance of the most important behavior modes of users, 
we compare the most important behavioral mode and the second
most important one in terms of their sizes. In Fig. \ref{cluster_1vs2}
we plot the size (i.e. number of vectors) distributions of the first
and the second behavioral modes under clustering threshold $0.2$
(solid lines) and $0.9$ (dotted lines) for USC users. We see that
there is a clear separation between the sizes of these two behavioral
modes. (i.e., the most dominant behavioral mode is much more important
than the second most important one for most users.) Different
clustering thresholds do not change the results much. In other words,
observations of the most dominant behavioral mode could reveal user
characteristic to a good extent for many users. Similar observations
also hold for Dartmouth users.

\begin{figure}

\centering
\includegraphics[width=2.5in]{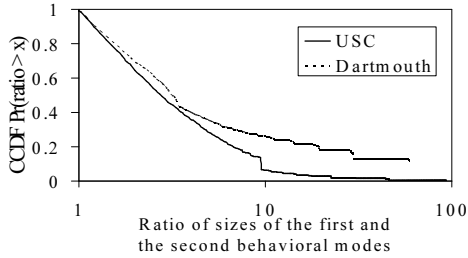} 

\caption{Complementary CDF for the ratio of the first behavioral mode
size to the second behavioral mode size. Note that the X-axis is in
log scale to make the graph more visible.}
\label{cluster_ratio}
\end{figure}

We show the distribution of the size ratio between the largest and the
second largest cluster in Fig. \ref{cluster_ratio}. Here we see for
USC and Dartmouth, respectively, $36\%$ and $31\%$ of users have the
two most important behavior modes with comparable sizes (i.e., with
size ratio smaller than $2.0$ - The second most important behavioral
mode is followed at least one half as often as the most important
behavioral mode). Hence looking at the most dominant cluster
exclusively could still be sometimes misleading and we might be ignoring
information about the user's detailed behavior. It is therefore
desirable to have a summary that takes not only the dominant
behavioral mode, but also the subsequent ones into account.

\subsection{Summarization Methods} \label{summary-proposals}

Now we investigate various ways to summarize the association
vectors, and then judge their quality based on a specific metric - the
{\it significance score}.

\noindent \textbf{\underline{Average of association vectors}}

This is the simplest way to calculate a summary. Averaging naturally
emphasizes  the dominant behavioral mode (as there are more
vectors in this mode). As users are not always online, the average
should include only the online days and ignore the zero vectors, defined as

$X_{onavg}$, denoted as
\begin{equation}
X_{onavg} = \frac{\sum_{i=1}^t  X_i}{\sum_{i=1}^t {\parallel X_i \parallel}_1},
\label{onavg}
\end{equation}
\noindent where ${\parallel X_i \parallel}_1$ is the L1 norm of vector $X_i$ (recall that for online days, the elements in association vectors sum to 1).

\noindent \textbf{\underline{Centroid of the first cluster}}

We observe for many users, the first behavioral mode is
dominant. Hence we can use
the centroid of vectors in the first {\it non-trivial} behavioral mode
(i.e., if the first behavioral mode is the cluster of zero vectors, we
take the second behavioral mode instead) as a summary. Formally,
\begin{equation}
X_{centroid1} = \frac{\sum_{X_i \in C_1} X_i}{\sum_{i=1}^t I(X_i \in C_1)},
\label{centroid1}
\end{equation}
where $C_1$ denotes the largest non-trivial behavioral mode for the
user and $I(\cdot)$ is the indicator function. Intuitively, it works well if the first behavioral mode
is dominant, but less so if there are multiple behavioral modes with
comparable importance for the user. We experiment with two different thresholds,
$0.5$ or $0.9$, to identify the dominant behavioral mode.

In order to quantitatively compare the quality of the summary techniques, we propose to measure the {\it significance score} of a summary vector with respect to a user by summing the projections of all association vectors on the summary vector, normalized by the online days of the user.
\begin{equation}
SIG(Y) = \frac{\sum_{i=1}^t  \vert X_i \cdot Y \vert}{\sum_{i=1}^t {\parallel X_i \parallel}_1},
\label{significance_score}
\end{equation}
\noindent where $Y$ is any summary vector. The physical interpretation
of the {\it significance score} is the percentage of power in
the association vectors $X_i$'s explained by the summary vector
$Y$. Following the definition, we calculate the average score of the
{\it significance} for $X_{onavg}$ and $X_{centroid1}$, and list
them in Table \ref{avg_significance}. We observe that the centroid
of the first cluster better explains the behavioral pattern of a given
user than the average, since averaging sometimes lead to a vector
that falls between the behavioral modes.

\begin{table}
\caption{The average significance score for various summaries of user association vectors}
\label{avg_significance}
\begin{center}
\begin{tabular}{|c||c|c|c|c|}
\hline
& \mr{$X_{onavg}$} & $X_{centroid1}$ & $X_{centroid1}$ & \mr{SVD} \\
& & threshold 0.5 & threshold 0.9 & \\
\hline
USC & 0.646 & 0.716 & 0.702 & 0.764 \\
\hline
Dartmouth & 0.690 & 0.757 & 0.747 & 0.789 \\
\hline

\end{tabular}

\end{center}
\end{table}

\noindent \textbf{\underline{Singular Value Decomposition}}

We revisit our definition of the {\it significance score} in
Eq. (\ref{significance_score}), and pose it as an optimization
question: Given the association vectors $X_i$'s, what is the best
possible summary vector $Y$ to maximize its significance?
Mathematically, we want the vector $Y$ to be
\begin{equation}
Y = \arg \max_{\Vert v \Vert =1} \sum_{i=1}^t \vert X_i \cdot v \vert.
\end{equation}
This is exactly the procedure to obtain the first singular vector if
we perform singular value decomposition (SVD) \cite{LP-norm} of the association
matrix $X$. In other words, if we want the summary vector $Y$
to capture the maximum possible power in the association vector
$X_i$'s, the optimal solution is to apply singular value decomposition
to extract the first singular vector. We apply this technique and
calculate the {\it significance score} in the last column in Table
\ref{avg_significance}. SVD provides the best summary.  Hence we use the SVD-based summary, and defer the
discussion of other summary techniques to section \ref{alternatives}.

\subsection{Interpreting Singular Value Decomposition}

In this subsection we explain other important properties of SVD
as applied to the association matrices.

\begin{figure}

\centering
\includegraphics[width=2.5in]{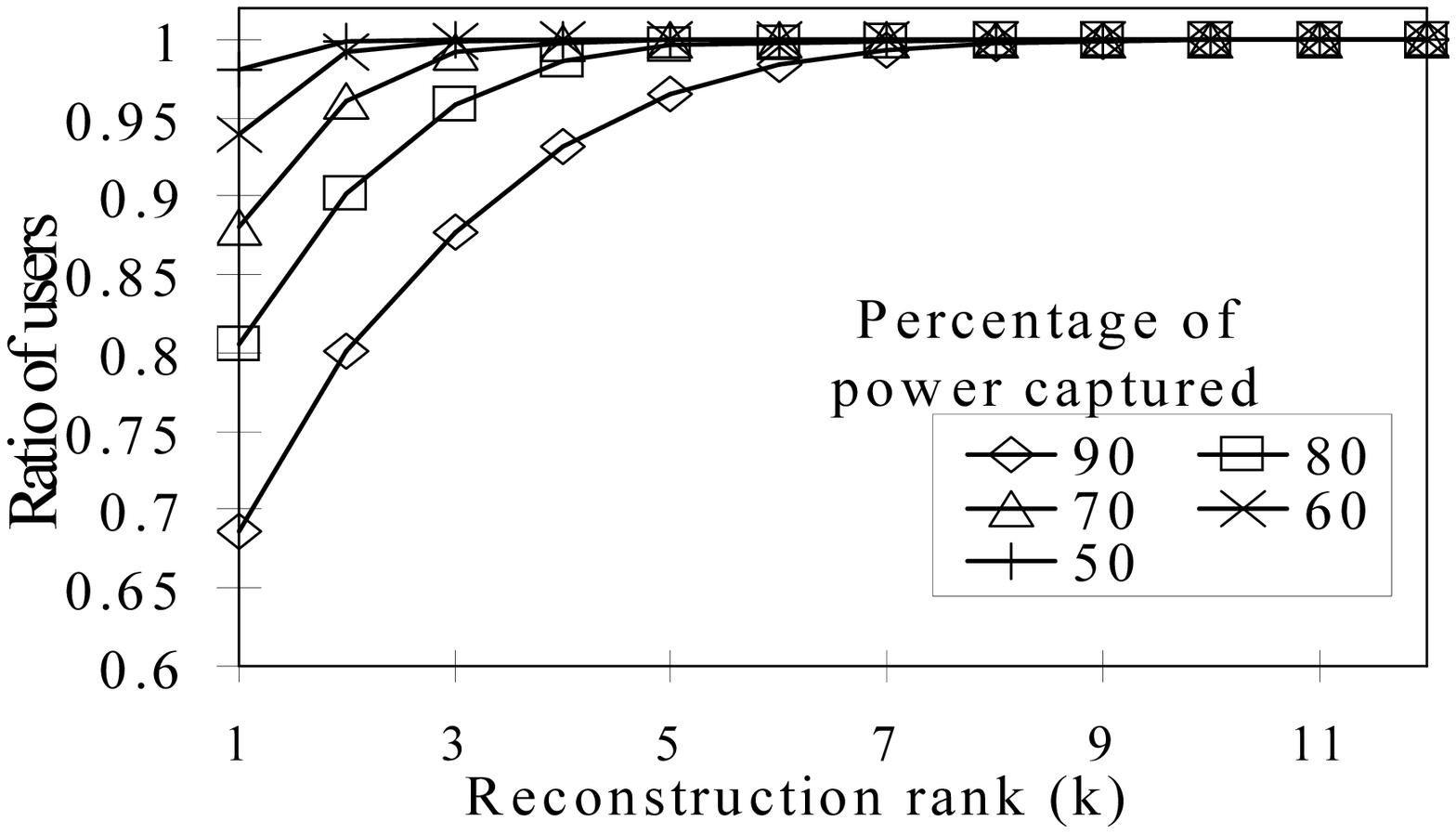} 

\footnotesize{(a)USC.}

\includegraphics[width=2.5in]{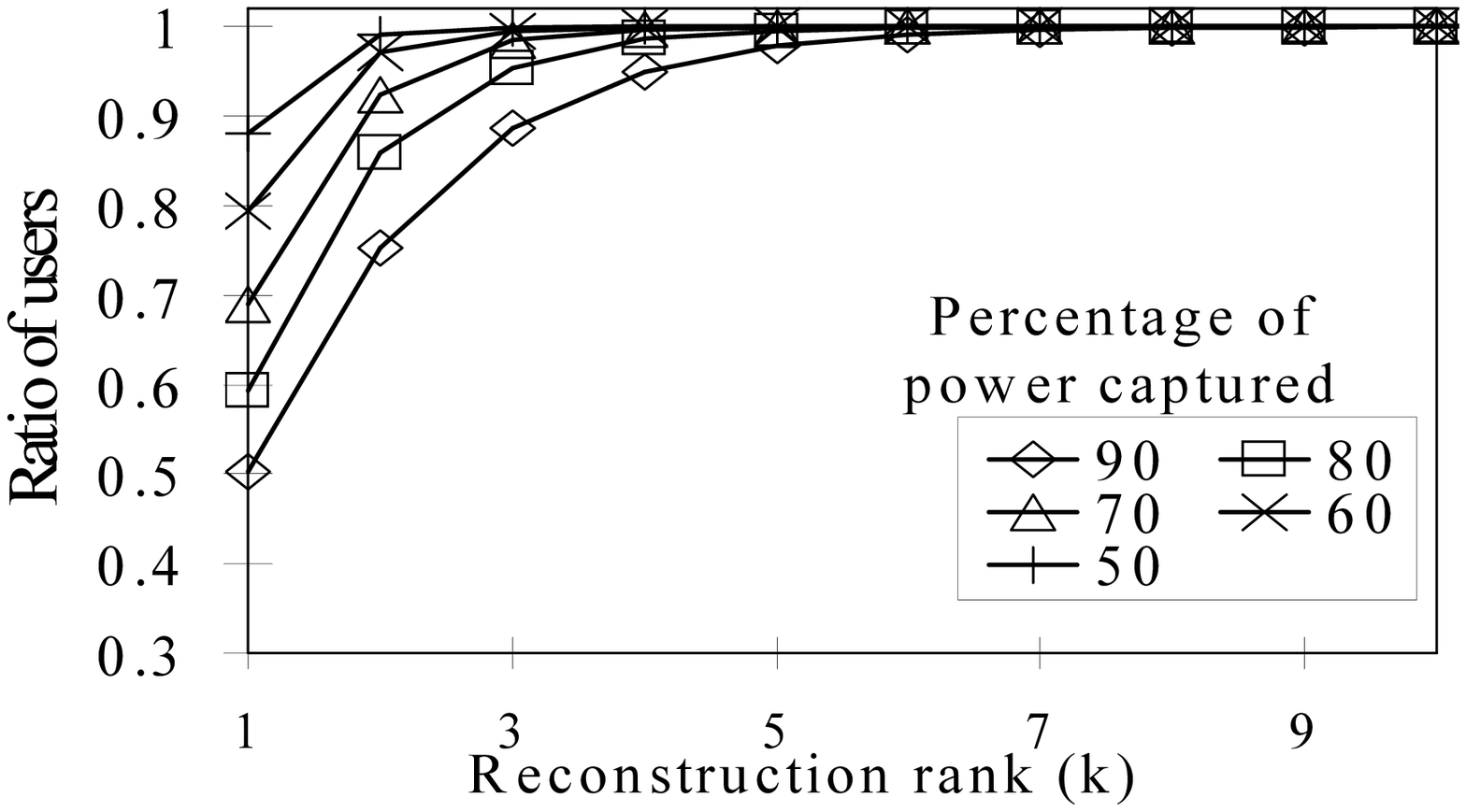} 

\footnotesize{(b)Dartmouth.}

\caption{Low association matrices dimensionality: A high target
percentage of power is captured with low rank reconstruction matrix
for many users.}

\label{percentile}
\end{figure}

From linear algebra \cite{LP-norm}, we know
that for any $t$-by-$n$ matrix $X$, it is possible to perform singular value
decomposition, such that
\begin{equation} X = U \cdot \Sigma \cdot V^T,
\label{SVD-eq}
\end{equation}
\noindent where $U$ is a $t$-by-$t$ matrix, $\Sigma$ is a $t$-by-$n$
matrix with $r$ non-zero entries on its main diagonal, and $V^T$ is an
$n$-by-$n$ matrix where the superscript $^{T}$ in $V^T$ indicates the
transpose operation to matrix $V$. $r$ is the rank of the original association
matrix $X$. The column vectors of the matrix $V$ are the
eigenvectors of the covariance matrix $X^T X$, and $\Sigma$ is a
diagonal matrix with the corresponding singular values to these
eigenvectors on its diagonal, denoted as $\sigma_1$, $\sigma_2$, ...,
$\sigma_r$. These singular values are ordered by their values
(i.e. $\sigma_1 \geq \sigma_2 \geq ... \geq \sigma_r$). We can re-write Eq. (\ref{SVD-eq})
in a different form:
\begin{equation} \tilde{X}_k = \sum^{k}_{i=1}  u_i \sigma_i v_{i}^T.
\label{recon}
\end{equation}
\noindent Here $u_i$'s and $v_i$'s are the column vectors of matrix
$U$ and $V$. They are used as the building blocks to reconstruct the
original matrix $X$. With this format, SVD can be viewed as a way to
decompose a matrix: It breaks the matrix $X$ into column vectors
$u_i$, $v_i$ and real numbers $\sigma_i$. If we retain all these
components (i.e., $k=rank(X)$), SVD is a lossless operation and the
matrix $X$ can be reconstructed accurately. However, in practical
application, SVD can be treated as a lossy compression and only the
important components are retained to give a rank-$k$ approximation of
matrix $X$. The percentage of power in the original matrix $X$
captured in the rank-$k$ reconstruction in Eq. (\ref{recon}) can be
calculated by
\begin{equation} \frac{\sum^{k}_{i=1} \sigma_i^2}{\sum^{Rank(X)}_{i=1} \sigma_i^2}.
\label{PC-percentage}
\end{equation}
For our data sets, users have much fewer behavioral modes
than the number of association vectors, and for most users the
dominant behavioral modes are much stronger than the others
(c.f. Fig. \ref{cluster_1vs2}). Hence we expect SVD to achieve great data
reduction on the association matrices. This is indeed the case, as we
show in Fig. \ref{percentile}: Most of the users have a high percentage
of power in association matrix $X$ explained by a relatively low-rank
reconstruction - For example, in the USC trace
(Fig. \ref{percentile}(a)), if we use a rank-$1$ reconstruction matrix, it
captures $50\%$ or more of power in the association matrices for more
than $98\%$ of users, and a rank-$3$ reconstruction is sufficient to
capture more than $50\%$ of power in association matrices for all
users. Even if we consider an extreme requirement, capturing $90\%$ of
power, it is achievable for $68\%$ of users using a rank-$1$
reconstruction matrix, and for more than $99\%$ of users using at most
a rank-$7$ reconstruction matrix. Similar observations can be made for
Dartmouth users (in Fig. \ref{percentile}(b)). For both campuses,
five components are sufficient to capture $90\%$ or more power
for most (i.e., more than $90\%$) of the users. This indicates
although users show multi-modal association pattern, for most users
the top behavioral modes are relatively much more important then the
remaining ones.

If a low-rank reconstruction of the association matrix is achievable,
it is natural to ask for the representative vectors for the behavioral modes
of a user. For this purpose, SVD can be viewed as a procedure to
obtain representative vectors that capture the most remaining power in
the matrix. Mathematically\footnote{SVD on matrix $X$ can be viewed
as calculating the eigenvalues and eigenvectors of the covariance
matrix, $X^T X$. This is also the procedure typically used to perform
Principal Component Analysis (PCA) for matrix $X$.},
\begin{align} \label{PCA-algebra}
\begin{split}
u_1 &= \arg \max_{\Vert u \Vert =1} \Vert  X \cdot u \Vert \\
u_k &= \arg \max_{\Vert u \Vert =1} \Vert ( X - \sum^{k-1}_{i=1} X u_i u_{i}' ) u \Vert \;\;\; \forall k \geq 2.
\end{split}
\end{align}
We can interpret the singular vectors, $u_j$'s, as the vectors that
describe the user's behavioral modes in decreasing order of importance
in the association matrix $X$, with its relative weight (or the
importance) quantified by $\sigma_j^2 / \sum^{r}_{i=1} \sigma_i^2$,
following Eq. (\ref{PC-percentage}). In the paper we refer to these
vectors as {\it eigen-behavior} vectors for the user.

The {\it eigen-behavior} vectors, $u_j$'s, are unit-length vectors. The
absolute values of entries in an {\it eigen-behavior} vector
quantify the relative importance of the locations in the user's
$j$-th behavioral mode. For example, suppose a given user visit
location $l$ almost exclusively, then in his first eigen-behavior vector, the
entry corresponds to location $l$ would carry a high value (i.e. close
to $1$), and the weight of the first eigen-behavior vector, $\sigma_1^2 /
\sum^{r}_{i=1} \sigma_i^2$, shall be high. With a set of {\it
eigen-behavior} vectors and their corresponding weights, we can
capture and quantify the relative importance of a user's behavioral modes.

There are several benefits of applying SVD to obtain the
summary as compared to other schemes: (1) SVD provides the optimal
summary that captures the most remaining power in the original matrix
with each additional component. (2) The components can be used to
reconstruct the original matrix, while the calculation of average or
centroid vectors are non-reversible. Thus SVD provides a
way to compress user association vectors and helps us save 
storage space. (3) Not only the most important
behavioral mode, but also the subsequent ones can be systematically
obtained with SVD, with a quantitative notion of their relative
importance.

\section{Clustering Users by Eigen-behavior vectors}  \label{feature}

In this section, we first define our novel distance measure based on the
 {\it eigen-behavior} vectors and then use it for clustering.
 
\subsection{Eigen-behavior Distance}  \label{feature-distance}

Suppose $u_i$'s and $v_j$'s are the eigen-behavior vectors of two users,
$i=1,...,r_u$ and $j=1,...,r_v$ where $r_u$ and $r_v$ are the ranks of
the corresponding association matrices. The
similarity between the two users can be calculated by the sum of pair-wise
inner products of their eigen-behavior vectors $u_i$'s and $v_j$'s, weighted by
$w_{u_i}$ and $w_{v_j}$\footnote{$w_{u_i}$ represents the
weight of the eigen-behavior vector $u_i$, calculated by $w_{u_i} =
\sigma_i^2 / \sum^{r_u}_{k=1} \sigma_k^2$. The weights $w_{u_i}$'s sum
up to $1$, and $w_{v_j}$'s are defined similarly.}. Our measure of similarity between two sets of eigen-behavior vectors, $U = \lbrace u_1,...,u_{r_u}\rbrace$ and $V = \lbrace
v_1,...,v_{r_v}\rbrace$, is defined as:
\begin{equation}
Sim(U,V) = \sum^{r_u}_{i=1} \sum^{r_v}_{j=1} w_{u_i}w_{v_j}\vert u_i \cdot v_j \vert.
\label{sim-index}
\end{equation}
Higher similarity index $Sim(U,V)$ indicates that the eigen-behavior vectors
$U$ and $V$ are more similar, and hence the corresponding users have
similar association patterns. We define the {\it eigen-behavior distance} between users
$U$ and $V$ as $D'(U,V)= 1-(Sim(U,V)+Sim(V,U))/2$.\footnote{We normalize the similarity indices from user $U$ to all other
users between $(0,1)$. Among all users, we find the user $K$ such
that $Sim(U,K) = \mathbf{max}_{\forall N} Sim(U,N)$. We than normalize
$Sim(U,V) = Sim(U,V)/Sim(U,K)$ for all users $V$.}

Using the {\it eigen-behavior distance} also reduces the computation
overhead.
If we use only the top-$5$ components (which captures more than $90\%$ power as in Fig. \ref{percentile}),
instead of going through $t$-by-$t$ pairs of original association
vectors as in section \ref{naive}, we reduce the distance calculation
to $5$-by-$5$ pairs. Since we have at least $61$ days in the traces,
this is at least a $(61/5)^2 \approx 148$ fold saving for all $N^2$ pair
of users. By paying the pre-processing (i.e., SVD for all $N$ users)
overhead of $O(Nt^2)$, we can reduce the distance calculation 
complexity from $O(N^2t^2)$ to $O(c \cdot N^2)$. Since
users follow repetitive trends in the association patterns, its total
{\it eigen-behavior} vectors would not grow with the number of time slots, $t$.
If we consider longer traces or association
vector representations in finer time scale, the reduction
can be even more significant.
In the following computations, we consider only the eigen-behavior vectors that
capture at least $0.1\%$ of total power.

\subsection{Significance of the Clusters} \label{feature-clustering}

\begin{figure}

\centering

\includegraphics[width=2.2in]{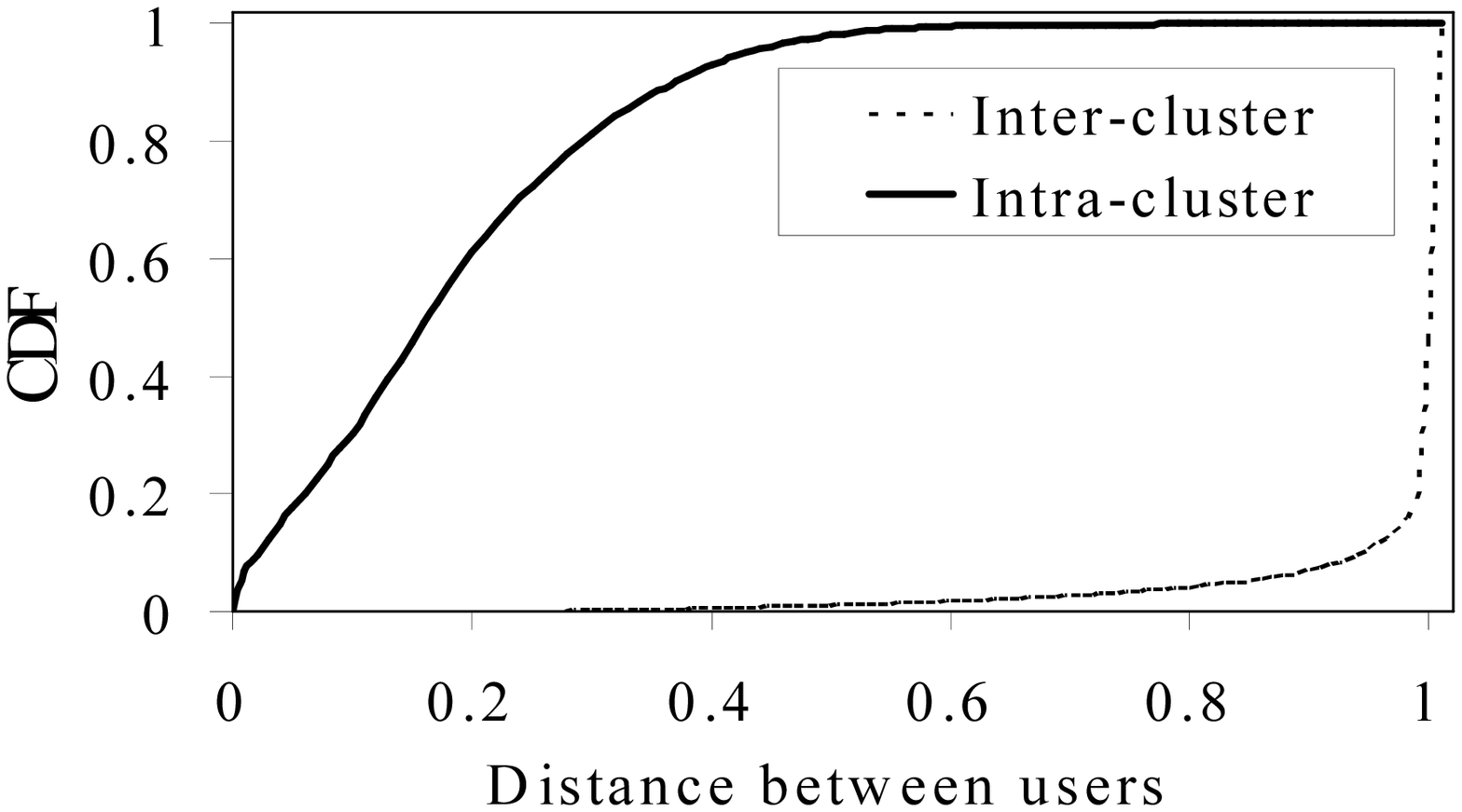}

\footnotesize{(a) USC.}

\includegraphics[width=2.2in]{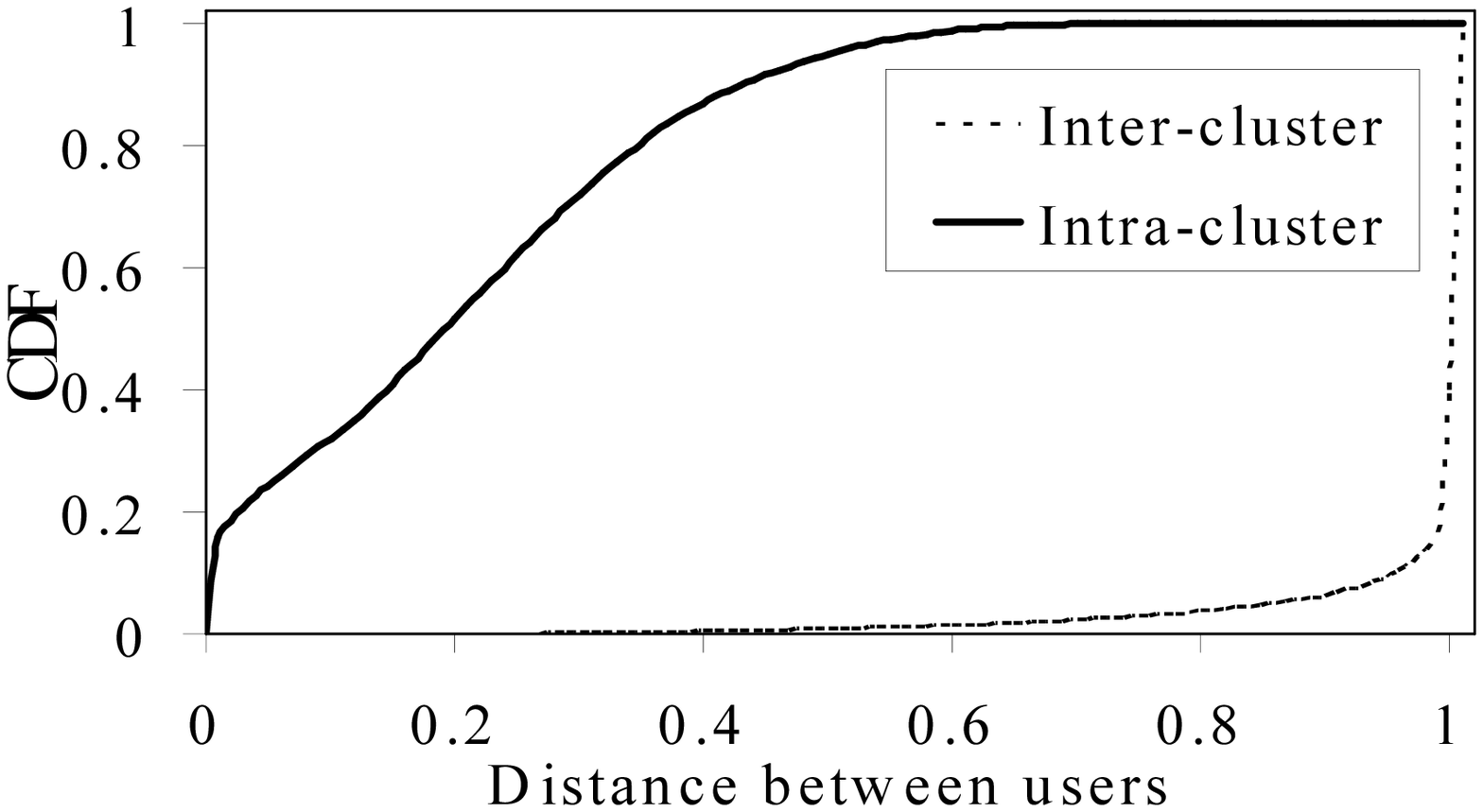} 

\footnotesize{(b) Dartmouth.}

\caption{Cumulative distribution function of distances for inter-cluster and intra-cluster user pairs (eigen-vector distance).}

\label{inter-intra-dist-feature}
\end{figure}

We cluster users based on eigen-behavior distance and again validate the results by plotting the intra-cluster and
inter-cluster distance distributions, when we consider $200$ clusters. With the eigen-behavior distance, for both
USC and Dartmouth traces, there is a better separation between the CDF
curves (Fig. \ref{inter-intra-dist-feature}) as compared to the results with the AMVD
distance (Fig \ref{inter-intra-dist-avgmind}), indicating a meaningful
clustering. This proves the eigen-behavior distance is a better
metric than the AMVD distance as it helps us to group users into
well-separated behavioral groups based on their WLAN association
preferences, for both campuses.

\begin{figure}
\begin{minipage}[t]{1.7in}
\centering

\includegraphics[width=1.7in]{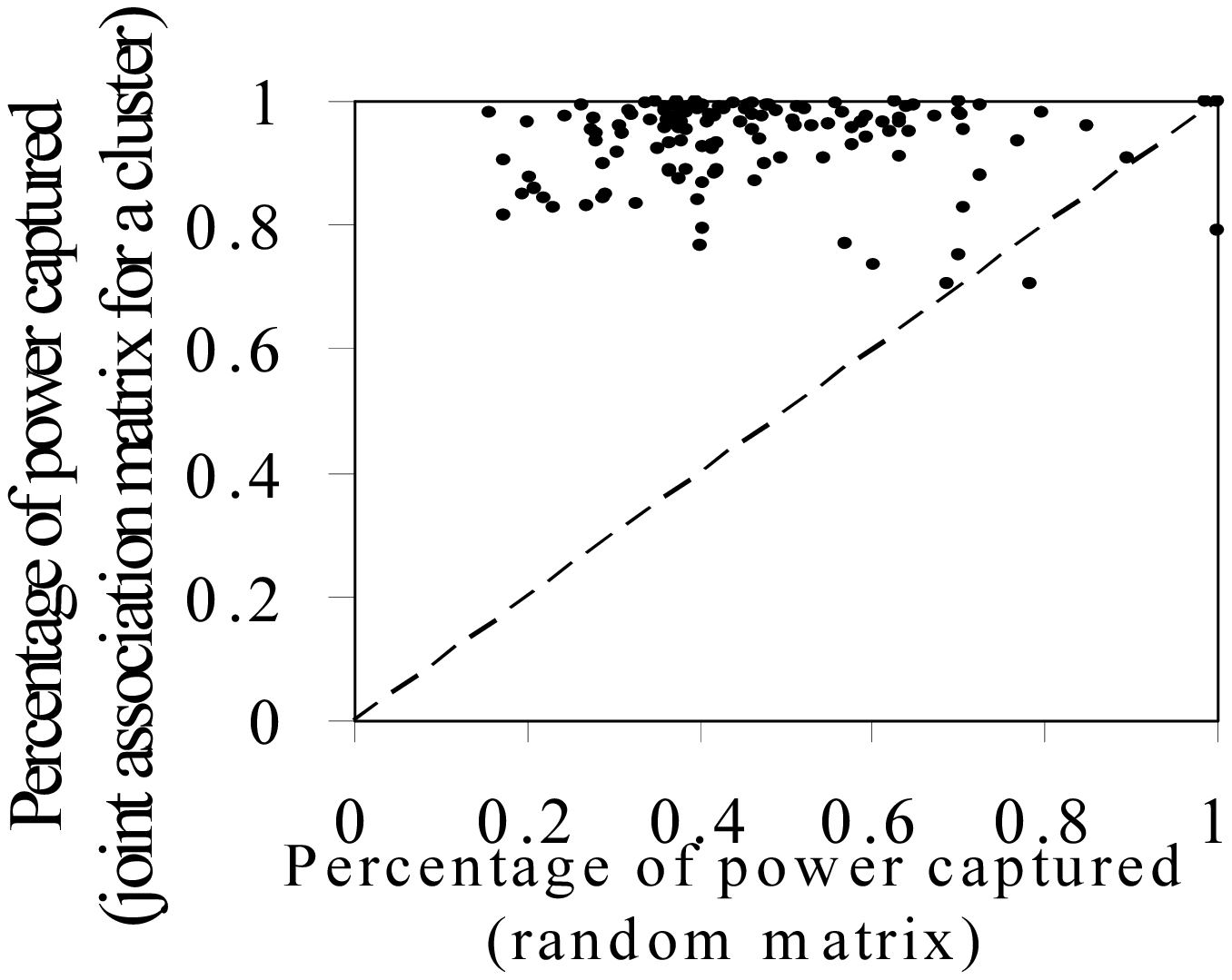} 

\footnotesize{(a) USC(129 clusters)}

\end{minipage}
\hfill
\begin{minipage}[t]{1.7in}
\centering

\includegraphics[width=1.7in]{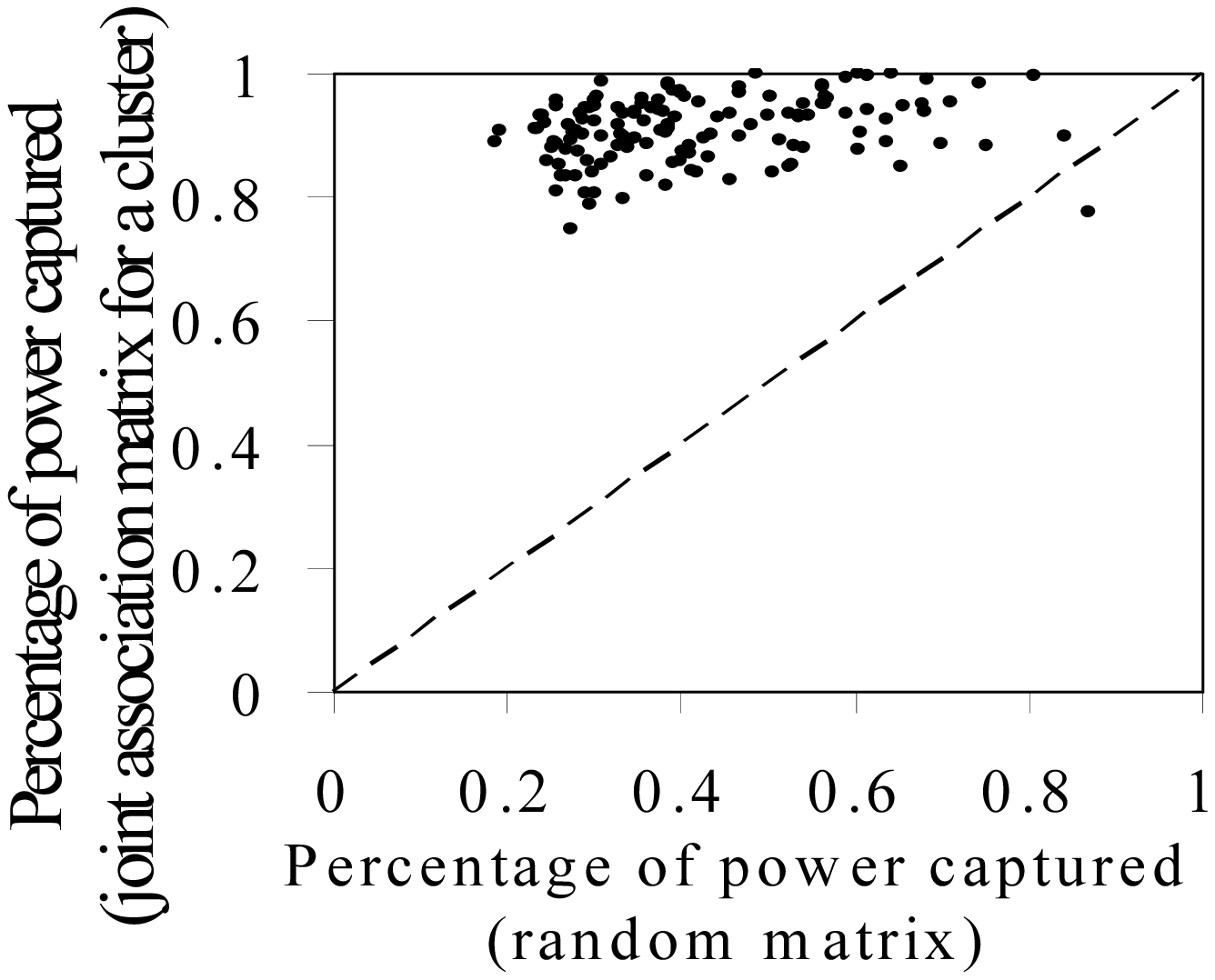} 

\footnotesize{(b) Dartmouth(136 clusters)}

\end{minipage}
\hfill
\caption{Scatter graph: Cumulative power captured in top four eigen-behavior vectors of random matrices (X) and joint association matrices formed by users in the same cluster (Y). Only clusters with $5$ or more members are included.}
\label{grouping-1stPC}
\end{figure}

We further validate whether the resulting clusters indeed capture users
with similar behavioral trends. We compose the {\it joint
association matrix} by concatenating the daily association vectors of
a cluster of $m$ similar users in a larger $mt$-by-$n$ matrix, where
$n$ is the number of locations and $t$ is the number of time
slots. When we perform SVD to the {\it joint association matrix}, the
top eigen-behavior vectors represent the dominant behavioral patterns within
the group. If the users in the group follow a coherent behavioral
trend, the percentage of power captured by the top eigen-behavior vectors
should be high. On the other hand, if association vectors of users
with different association trends are put in one {\it joint
association matrix}, the percentage of power captured by its top
eigen-behavior vectors should be much lower.
Among all clusters, we pick those with more than five users, and
compare the cumulative power captured by the
top four eigen-behavior vectors of these clusters with 
random clusters of the same size in scatter graphs, Fig. \ref{grouping-1stPC}.
Clearly, most the dots are well
above the $45$-degree line for both campuses. This indicates the users in the
same cluster follow a much stronger coherent behavioral trend than
randomly picked users, pointing to the significance of our clustering results.

We would also
like to see if each cluster from the population shows a distinct
behavioral pattern.  To quantify this, we obtain the first
eigen-behavior vector from each group and calculate its {\it significance
score}, defined in Eq. (\ref{significance_score}), for all the groups.
The results confirm with our goal of identifying groups following
different behavioral trend: For the USC trace, the first
eigen-behavior vectors obtained from the {\it joint association matrices} have
an average {\it significance score} of $0.779$ for their own clusters and
an average score of $0.005$ for other clusters, indicating the
dominant behavioral trends from each cluster is distinct. The
corresponding numbers for the Dartmouth trace are $0.727$ and $0.004$,
respectively.

We conclude that we have designed a distance
metric that effectively partitions users into groups based on behavioral
patterns. In addition, these clusters are unique with respect to
behavioral trends.  


\section{Interpretation of the Clustering Results} \label{interpretation}

In this section we analyze and interpret the results of clustering for both
university campuses from social perspective. 

\begin{figure}

\centering
\includegraphics[width=2.5in]{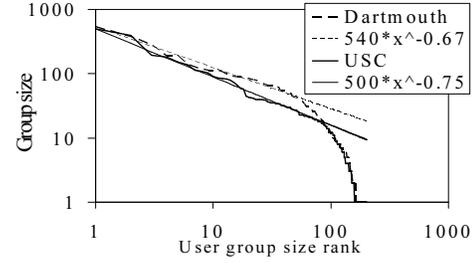} 

\caption{Rank plot (group size ranking v.s. group size) in log-log scale. User group size follows a power-law distribution.}
\label{group_size}
\end{figure}

First we analyze the group size distribution, as
shown in Fig. \ref{group_size}.  We observe the distributions of group
sizes are highly-skewed for both campuses. There are dominant
behavioral groups that many users follow: the largest groups in the
campuses include $504$ and $546$ members, out of the population of
$5000$ for USC and $6582$ for Dartmouth, respectively. The ten largest
groups combined account for $39\%$ and $33\%$ of the total population,
respectively.  On the other hand, there are also many small groups,
or even singletons, for both populations: out of the $200$ clusters,
there are $68$ and $57$ of them with less than five members,
respectively, and in both campuses about half of the groups have less
than $10$ members. More interestingly, we observe that besides these
small clusters, the distribution of the cluster size follows a
power-law distribution. In Fig. \ref{group_size}, we plot the
straight lines that illustrate the best power-law fits. The slopes for
these lines are $-0.67$ for Dartmouth and $-0.75$ for USC, respectively.
The power law distribution of group sizes may be related to the skewed popularity
of locations on campuses - it has been shown that the number of patrons to various locations
differ significantly\cite{Dart-trace}.  However, the link between the
distributions of number of patrons and the distribution of group sizes
is not direct. While the most-visited locations on both campuses
easily attract thousands of patrons, these people are broken into
different behavioral groups depending on their association
preferences.

We now study the detailed
behaviors of each cluster by using the eigen-behavior vectors
and their relative weights to understand the detailed preferences of
the groups. We discover for most of the groups, their top
eigen-behavior vectors dominate i.e. the
contribution of the second-most important location is almost invisible
in the first eigen-behavior vector. Similar relationship holds between the
second-most important location and the third-most important one, and
so on. Hence the association behavior of the group can be described by
a sequence of locations of decreasing importance with a clear
ordering. This observation matches with the current status of WLAN
usage: people tend to access WLAN at only a limited number of
locations, and the preference of visiting locations is skewed
\cite{winmee-indi}.  For such users, its most visited locations
might be sufficient to classify them.

Most large user clusters belong to the fore-mentioned case. The
largest clusters on both campuses include the library visitors, as expected,
since libraries are still the most visited area on university
campuses. For the USC campus, the largest user cluster visits the
library (the first eigen-behavior vector has a single high-value entry
corresponding to the library, and this eigen-behavior vector captures $83\%$
of the power in the joint association matrix for the group), followed
by a couple locations around the Law school ($4.45\%$) and the school
of Communication ($4.5\%$), both are popular locations on campus.  For
the Dartmouth campus, the largest user cluster visits LibBldg2
($72.85\%$), followed by LibBldg1 ($5.13\%$), SocBldg1 ($3.56\%$), and
LibBldg3 ($1.93\%$).  It seems this group consists of library patrons
who mainly move about the public area on the campus and access the
WLAN from these locations.

While libraries are popular WLAN hot spots, we also discover many user
clusters that rarely visit these locations.  The second largest
cluster for USC consists of users visiting mostly the Law school
($89.73\%$ of power), school of accounting ($6.37\%$), and a couple of
locations close to the Law school ($0.59\%$). For Dartmouth, the
second largest cluster visits AcadBldg18 ($56.38\%$), AcadBldg6
($13.4\%$), ResBldg83 ($10.15\%$), AcadBldg31 ($3.5\%$), AcadBldg7
($3.12\%$), which seems to be a group of students going to classes at
multiple academic buildings. We have also observed various clusters
featured different dorms and classrooms as their most visited location
from both campuses.

On the other hand, we have also discovered groups with multiple
high-value entries in its top eigen-behavior vectors from both campuses.  One
prominent example from USC trace consists of $32$ users, who visit
buildings VKC and THH, two major classrooms on the USC campus.  The top two
eigen-behavior vectors of the cluster both consist of two high-value entries
corresponding to these two buildings\footnote{One of the
eigen-behavior vectors has positive values for both entries, and the other
has one positive and one negative value, in order to adjust the ratio
between these two locations in the association vectors.}, and they
capture $63.14\%$ of power in the joint association matrix. This
cluster consists of users who visit these two locations with similar
tendency, according to the eigen-behavior vectors, and such distinct
behavioral trend exists for $32$ users in the population. This cluster
is a good example to show why it is not sufficient to merely use the
most dominant behavioral mode (or the most-visited location) of a user
to classify it. If the centroid of the dominant behavioral mode (i.e.,
Eq (\ref{centroid1})) is used to classify users, the behavioral trend
of visiting multiple locations with similar tendency will not be
revealed.  Instead, among the $32$ users, $13$ are classified with
others who visit VKC frequently, $10$ are classified with those who
visit THH frequently, and the rest are put into various groups. As
portable wireless devices gain popularity, we expect to see more users
displaying diverse behavioral trends in terms of network usage. To fully
capture such behavior, averaging-based summary is not sufficient, and
this is where SVD shows its strength the most.

Interestingly, we also discover many small clusters with unique
behavioral patterns that deviate from the "main stream" users in both
traces. For example, in the USC trace, there is a small cluster of
eight users who visit exclusively a fraternity house.  Probably these
are the people who live there. In the Dartmouth trace, there is a
cluster of eight users who visit mostly athletic buildings (AthBldg5
($90.9\%$), AthBldg10 ($4.62\%$), AthBldg2 ($3.14\%$), AthBldg3
($0.8\%$), and ResBldg26 ($0.54\%$)).  These are probably either
athletes or management staffs of the athlete facility. Such findings
substantiate our motivation of the study: as the wireless technology
prevails, we can expect users to display diverse behavioral patterns
that reflect to their personal preferences, and it is important to
capture such behavioral trend and quantify its significance.

To sum up, we have demonstrated  a systematic way to
identify distinct behavioral groups within on-campus populations, by using clustering based on
association features obtained from large-scale wireless network traces.
The method and findings are useful for various applications, as we show in the subsequent section with a case study of profile-based casting, and discuss other potential applications in the next section.

\section{Case Study: Mobility-profile-based Casting Protocol} \label{MPcast}

\subsection{Preliminaries}

Delay tolerant networks (DTNs)\cite{DTN} are networks characterized by sparse,
time-varying connectivity, in which end-to-end spatial paths from
source to destination nodes are often not available.  Messages are
stored in intermediate nodes and moved across the network with nodal
mobility. One particular important decision to make for nodes in DTN
is whether to forward a packet to other nodes they encounter (i.e.,
move into the radio range) with.  Such decisions have implications on
many aspects of how efficiently the routing strategies work, such as
delay, overhead, and message delivery rate.

\subsection{A Mobility {\it PROFILE-CAST}ing Protocol}

In the paper we consider the scenario where the message sender is
interested in forwarding messages to users with a similar mobility
profile.  For example,
a student loses a wallet and wishes to send an announcement to
other fellow students who visit similar places often as he does to
look for it. Or, for location specific announcements such as power
shutdown in parts of campus directed only to patrons of the specific
area. Note that this application is different from {\it geo-casting},
which targets at the nodes {\it currently} within a geographical
region as the receivers. Our target receivers are nodes with a certain
mobility profile, {\it regardless} of their actual locations at the
time the message is sent. To enable such {\it profile-based casting}
services, it is important to have a descriptive representation for
user behavioral profiles and a measure of the similarity between
users, to guide the message forwarding decisions.

In previous sections, we analyze two large scale WLAN user
association traces\cite{MobiLib-data, Dart-movement-data} and
represent user mobility in the form of {\it daily association
vector} to classify the whole user population into distinct
behavioral groups with unsupervised learning techniques. In this section,
we take these behavioral groups as the targets for mobility
profile-based casting.

The details of our {\it similarity-based} profile-based casting is as
follows: Users in the network are not aware of the centralized
decision of user grouping based on similarity in their
mobility. Instead, when two users meet each other, they exchange the
mobility profiles (i.e., vectors with their relative importance
(weights)) of their previous behavioral patterns and decide whether they are similar
at the spot, according to Eq. (\ref{sim-index}).
If the similarity index is larger
than a threshold, they exchange the message.
Note this decision is solely local, involving only the two encountered nodes.
The philosophy behind the protocol is, if each node delivers the
message only to others with high similarity in mobility profile, the
propagation of the message copies will be scoped.

\subsection{Evaluation and Comparison}

We compare the performances of following schemes with the {\it
similarity-based} protocol : (1) {\it Flooding}: The nodes in the
network are all oblivious to user mobility profiles and blindly
send out copies of the message to nodes who have not received it
yet. This scheme is also known as {\it epidemic routing}
\cite{epidemic}.  (2) {\it Centralized}: In this ideal scenario, all
nodes acquire the centralized knowledge of the behavioral group
membership, and only propagate the message to others if they are in
the same group. The message will never propagates to an unintended
receiver.  (3){\it Random-transmission (RTx)}: The current message
holder sends the message to another node randomly with probability $p$
when they encounter, and never transmits again (i.e., only the node
who last received the message will transmit in the future).  Loops are
avoided by not sending to the nodes who have seen the same message
before. This process continues until a pre-set hop limit is reached.

We utilize the USC trace \cite{MobiLib-data} to study the message
transmission schemes discussed above {\it empirically}. We use the
trace for user mobility and assume that two nodes are able to
communicate when they are associated to the same access point.  Note
that the WLAN infrastructure is merely used to collect user location
information, and the messages can be transferred only between the
users without using the infrastructure, as in
\cite{M-space-routing}. We split the WLAN trace into two halves. The
first half of the trace is used to determine the grouping of users
based on their mobility and we choose the number of
clusters to be $200$. Then we evaluate the group-casting protocol
performances using the second half of the same trace. For each group
with more than $5$ members, we randomly pick $20\%$ of the members as
the source nodes sending out a {\em one-shot message} to all other
members in the same group.

The performance metrics used are as follows: (1) {\it Delivery ratio}:
The number of nodes receiving the message over the number of intended
receivers.  (2) {\it Delay}: The average time taken for a scheme to
deliver the messages to recipient nodes.  (3) {\it Overhead}: The
total number of transmissions involved in the process of message
delivery.

\begin{figure}
\centering
\includegraphics[width=2.8in]{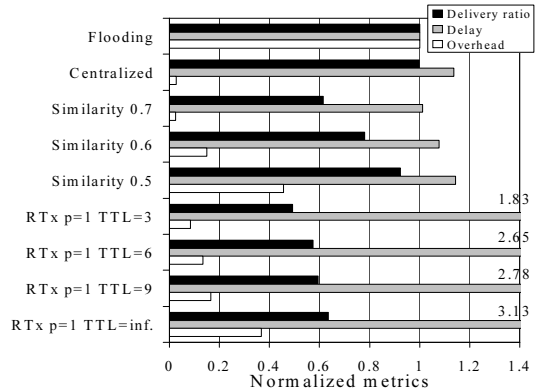} 

\caption{Relative performance metrics of the group-casting schemes
normalized to the performance of {\it flooding}.}

\label{protocol-metrics}
\end{figure}

We choose {\it flooding (i.e., epidemic routing)} as the baseline for
our evaluation and show the relative performance of the other
group-casting protocols relative to that of epidemic routing in
Fig. \ref{protocol-metrics}. In the graph we see that {\it flooding}
has the lowest delay and the highest delivery ratio as it utilizes all
the available encounters to propagate the message. However, it also
incurs significant overhead. The average delay, which is the lowest
possible under the given encounter patterns, is in the order of days
($3.56$ days in this particular case).  Group-casting based on {\it
centralized} clustering information, the ideal scenario, shows great
promise of behavior-aware protocols, as it significantly reduces the
overhead while maintains almost perfect delivery ratio, with a little
extra delay. However, it is not realistic to assume such centralized
knowledge.

For the {\it similarity-based} protocol, its aggressiveness can be
fine-tuned with the forwarding threshold of the similarity
index. Experiment results show a significant reduction of overhead
(only $2.5\%$ of {\it flooding}) at the cost of delivery ratio if we
set a high threshold such as $0.7$ (i.e., sending almost exclusively
within the same group).  Setting a low threshold (e.g., $0.5$) leads
to better delivery ratio ($92\%$ of {\it flooding}) but
still cuts the overhead to $45\%$ of {\it flooding}.  For the {\it
RTx} protocol, although the overhead can be controlled with the
hop-limit (which we set as $TTL$ times of the group size), we see that
the delivery ratio is lower than that of the {\it similarity-based}
protocol with comparable overhead (comparing {\it similarity $0.6$}
with {\it RTx $TTL=9$}, the former has $30\%$ higher delivery ratio
than the later) because in many cases the message is transmitted to
some node out of the desired group and there is no knowledge to direct
its propagation. Further more, the average delay for the delivered
messages is much longer than in the other protocols where multiple
copies of message propagate in the network. In addition, we try the
{\it RTx} protocol with various $p$ and $TTL$ values and find it is
not as flexible as the {\it similarity-based} protocol in which the
parameters can be tuned to trade overhead for better delivery ratio.

\section{Discussions} \label{disc}

\subsection{Potential Applications} \label{apps}

The insights of user grouping obtained from our
analysis can be applied in many other ways. We discuss some of these
application in this section, including
(1) network management, (2) user modeling, (3) behavior-aware services.

\textbf{\underline{Network Management}} Our analysis provides a different view of network management.  WLAN management and planning could be done by monitoring the activities of  individual APs  in order to identify the busy ones.
From the clustering technique, the manager can identify user groups and the relative importance of locations to each group. Such information can be helpful in terms of load prediction and planning. For example, if the business school is going to expand, by checking the behavioral groups of business school students, it is possible to predict its impact on the load of different parts of the WLAN.  For better understanding, one may also observe the change in the group structure with time and across semesters.

The SVD techniques detailed in this paper also provides a succinct way to express the {\it normal} behavior of a given user. Once the {\it norm} of the user's behavioral pattern is established, the system administrator could use the knowledge for behavioral abnormality detection. Obvious deviation in current behavior from the norm could be due to an impersonation attack or theft of the device, and should be brought to the administrator's attention depending on the policy.

\textbf{\underline{User Modeling}} Results from the clusters of users could help us to propose more realistic models for WLAN users, which is a challenge and a necessity  for evaluating network protocols. Although mobility models with groups of user is not a new idea\cite{RPGM}, there has been little work in realistic models based on groups.
Our decomposition approach provides two pieces of important information: (1) the distribution of group sizes follows a power-law distribution and (2) the detailed eigen-behavior vectors of the groups. With such information, one can set up generative model with the group sizes and the weights for frequently visited locations (e.g., its {\it communities}\cite{TVCM}) properly to evaluate their impacts on the network.

\textbf{\underline{Behavior-aware Services}} In future,  we expect the wireless devices have to be very portable and personalized. Hence, the services provided could be highly personalized, or at least customized based on the {\it interest groups}. Our method would facilitate to identify the dominant groups. The case study in the previous section shows that we can utilize mobility profile as a basis for such grouping and guidance of message delivery.
Certainly, different {\it representations} of users (e.g. hobbies, interests) that fit into the context might also be utilized rather, but our method would still be applicable. Furthermore, the service providers could assign a {\it target behavioral vector} to describe the property of target users, and the user devices could easily determine potential customers using a {\it significance score} (i.e., Eq. (\ref{significance_score})) to compare its eigen-behavior vector to the target behavioral vector. We refer to this scenario as interest-based grouping and profile-casting, and it is our future work.

In addition to clustering, the eigen-behavior vectors could also provide an efficient mechanism for users to exchange their behavioral features in order to {\em make new friends}.  Such social profiles could be applied in applications in social networking, such as behavior pattern oriented matching.

As large-scale city-wide WLAN deployments become commonplace, solutions to issues in management, service
design, and protocol validation could immensely benefit from insight into the behavioral patterns of the users or the {\em society}. We believe that our framework will be able to  provide a the behavioral patterns and help find solutions to several problems ranging from wireless network management to understanding basic social behavior of users armed with mobile devices in large WLANs.

\subsection{Alternative Representations and Metrics} \label{alternatives}

We have evaluated our {\it TRACE} approach extensively with other distance metrics
and representations of the data. Due space constraints, we only briefly discuss them here. Please refer to Appendix A for more details.

We design distance metrics with other types summaries in section \ref{summary-proposals}, and
they lead to user partitions similar to that from the SVD-based summary,
since in current WLANs, most users have a dominant behavioral mode so
simple summaries suffice to capture the trends. However, SVD is
able to discover repetitive trends when users have a complicated pattern of visiting association points,
while other methods cannot, as we argue in section \ref{interpretation}.

We also consider several other representations. Without normalization (i.e., the entries in the association vectors represent the absolute duration of association), the most active users are classified similarly as in the case of normalized representation, but the less active users fall in different clusters due to their sporadic usage of WLAN. Our idea is to view a user's behavior based on the fraction of time she spends at a location. We also explore the clustering by using both finer location granularity ( i.e., use each AP as a unique location) as well as finer time granularity (i.e. one association vector for each three-hour period). The resulted clusterings are similar to what we obtain, indicating the time-scale of daily vectors with per-building location granularity provide sufficient information.


On a different note, it may be of independent interest to use our representations in other domains in
different type of networks. For example, in encounter-based networks \cite{enc-net}, a representation
of encounter probability or duration would be appropriate. We plan to investigate this in our future work.

\section{Related Work} \label{rw}


As wireless networks gain popularity, it is extremely important to
understand its characteristics realistically. Along this line, there
have been great efforts to collect traces from WLAN users.
For examples, see \cite{Dart-trace, winmee-indi, MIT-trace, UCSD-trace}.
Many more traces have been made available
through efforts to build libraries of measurement traces
\cite{CRAWDAD-web, MobiLib-web}.

Although user association pattern has been one major focus in studies
about WLANs, for most previous works the focus is either on
aggregated statistics or on association models for individual user.
For the aggregate statistics, the current operating status of WLANs
is studied extensively, including user association preferences and durations,
mobility and hand-off, among others.
Henderson et al. focuses on the comparison of the same campus during
different time periods\cite{Dart-trace}. Balazinska et al. emphasizes on the mobility of corporate
WLAN users\cite{MIT-trace}. \cite{UCSD-trace} is a study specific to PDA users and their
mobility. These traces are compared in \cite{winmee-indi} based on
aggregate statistics.
For most of the modeling works, the focus is to obtain particular statistics about users
and to establish a model based on these quantities. In most of these modeling
efforts, the users are considered as
independent samples from a uniform population.
In \cite{NCCH-trace} the user association durations are modeled
by BiPareto distributions.
In \cite{WLAN-model} the authors match user session lengths and
hand-off probabilities between APs to generate a mobility model.
In \cite{ModelT, ModelT++}, the authors further cluster the locations (i.e., AP)
based on the number of user hand-off between them to generate
a hierarchy in user hand-off model. Hsu et al. explicitly models periodicity
of users visiting their favorite locations\cite{TVCM}.
There are hardly any studies on understanding the relationships
between users in the literature. The only excpetion we are aware of is perhaps \cite{classify-Dart}, where Kim et al. look
into the range of movement of users, and classify users based on the
periodicity of the movement range.
We provide a further
step towards this understanding by classifying users into groups of
similar behavior. This provides a different and important perspective
to understand user association patterns.

There are several papers in the literature that also use clustering
techniques. One with a similar goal to ours is \cite{classify-Dart},
which classifies users based on a different {\it representation}. In their paper,
users are classified based on the dominant periods in
their movement (e.g. those who display strong daily or weekly movement
patterns) and their longest movement ranges, but not based the location preferences. Hence the results
have different interpretations to ours.
In \cite{UW-places} the authors apply clustering
technique to the trace of location coordinates of a user to discover
significant places for the user, but they have not focused on
classifying users.

The technique we utilize to obtain association features from users,
singular value decomposition\cite{LP-norm}, is widely-applied to
discover linear trends in large data sets. It is closely related to
principal component analysis \cite{PCA-book}.  In \cite{eigenflow},
the authors utilized PCA to decompose the traffic flow matrices for
ISP networks and understand the major trends in the traffic. Our
application of SVD to individual user association matrices is similar
in spirit to their work. Note that it is typical for people to follow
dominant routines in lives, hence we expect the SVD approach to be
applicable to various human behavioral data sets.  In
\cite{reality-mining}, the authors also use PCA to discover trends in
a cellphone user group, which is similar to our analysis on individual
users. In this paper, in addition to analyzing much larger data sets,
we further compare user similarity and define distance metrics to
classify wireless network users into groups with robust validation.
Note that in order to make the eigen-behavior vectors obtained from all users
comparable, we need to keep the origin fixed among all association
matrices. Hence we adopt a variant, called {\em uncentered PCA}
\cite{PCA-book} where the mean of each dimension is not subtracted.
It has been used to study the diversity of species at various
sites\cite{uncentered-PCA}.

\section{Conclusion} \label{conclusion}

In this paper, we classify groups of WLAN users based on the trends in
their association patterns in two major university campuses by
leveraging clustering techniques and our systematic {\it TRACE} approach.
We design a novel distance metric between users based on the
similarity of their {\it eigen-behavior} vectors, obtained through singular
value decomposition (SVD) of the association matrices. SVD is the
optimal way to capture underlying trends in the data set, and we have
shown although many (at least $60\%$) users display multi-modal
behavioral modes, SVD is able to capture at least $90\%$ of power in
association matrices for most users with at most five components. This
also leads to space and time efficient computations.

The eigen-behavior distance leads to a meaningful partition of
users. We establish that WLAN users on university campuses form a
diverse community, which includes hundreds of distinct behavioral
groups in terms of association patterns. The size of the groups
follows a power-law distribution on both campuses.  While the large
groups account for a major part of the population (the top ten groups
account for at least $33\%$ of population in our data sets), there
exist many small groups with unique association patterns. In spite
of the very different location and demography  of the two university campuses, it is
surprising to find out the qualitative commonalities of the user
behavior trends.

While distance metrics based on simple summaries (e.g., average or
centroid of the dominant behavior mode) suffice for most current
WLAN users, our study indicates that SVD is capable to capture user association
trends in complex situations e.g. when users visit several distinct locations
on different time slots (e.g. days).
As personalized wireless devices become more popular, WLANs become ubiquitous and
their powerful combination impacts our daily lives, a
powerful tool to understand the user behavior is essential. Such
understanding could lead to better network management, user behavior
modeling, or even behavior-aware protocols and applications.

\section*{Appendix A. Alternative Methods} \label{alternatives}

Besides the {\it normalized association vectors} and {\it eigen-behavior distance} obtained through SVD, there are many other potential representations of user association behavior and distance metrics. In this section we discuss about these alternatives and some results we obtain with those.

\subsection{Various Distance Metrics} \label{alternative-metric}

\begin{figure}
\begin{minipage}[t]{1.6in}
\centering

\includegraphics[width=1.6in]{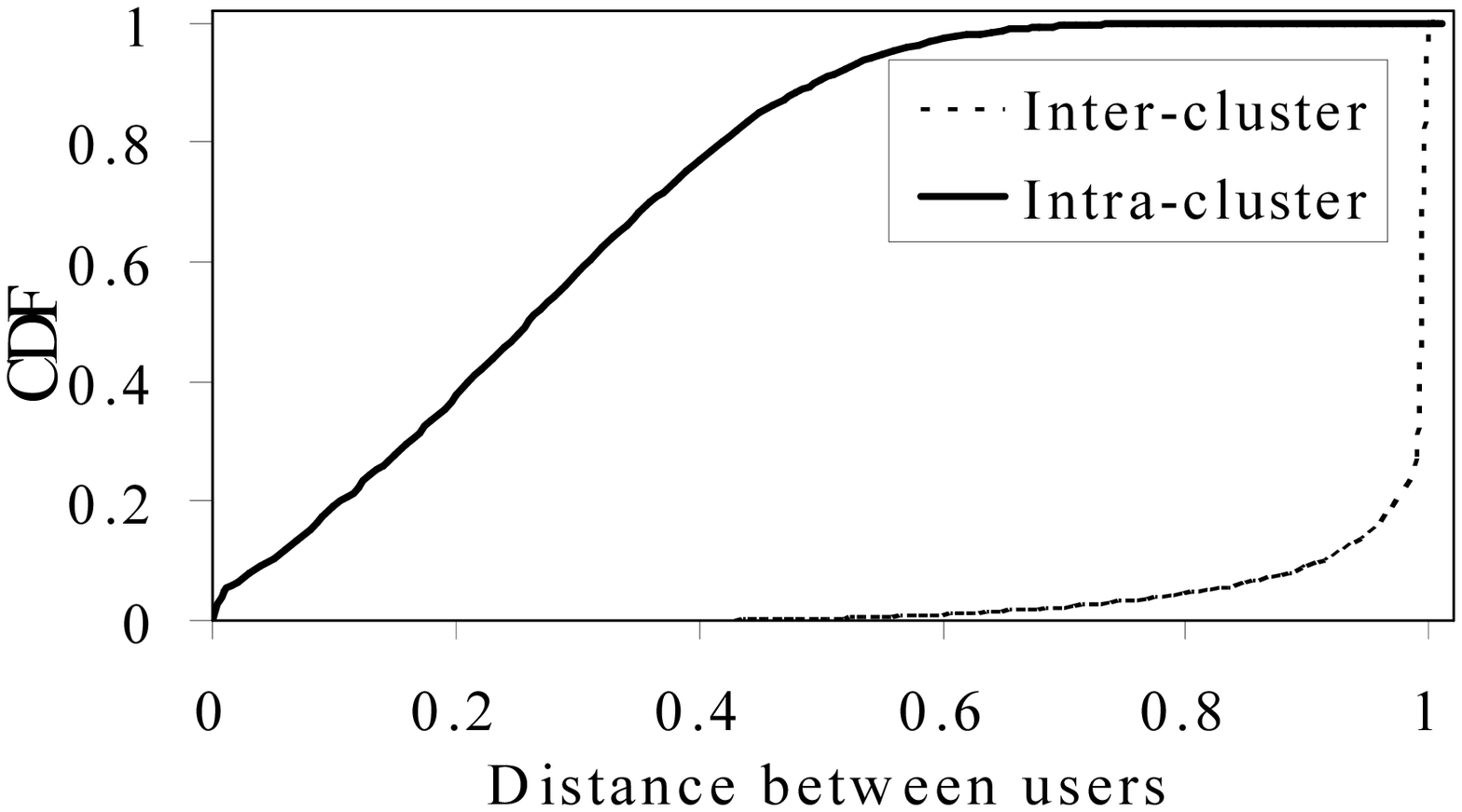}

\footnotesize{(a) USC, distance between average of association vectors.}
\hfill
\end{minipage}
\begin{minipage}[t]{1.6in}
\centering
\includegraphics[width=1.6in]{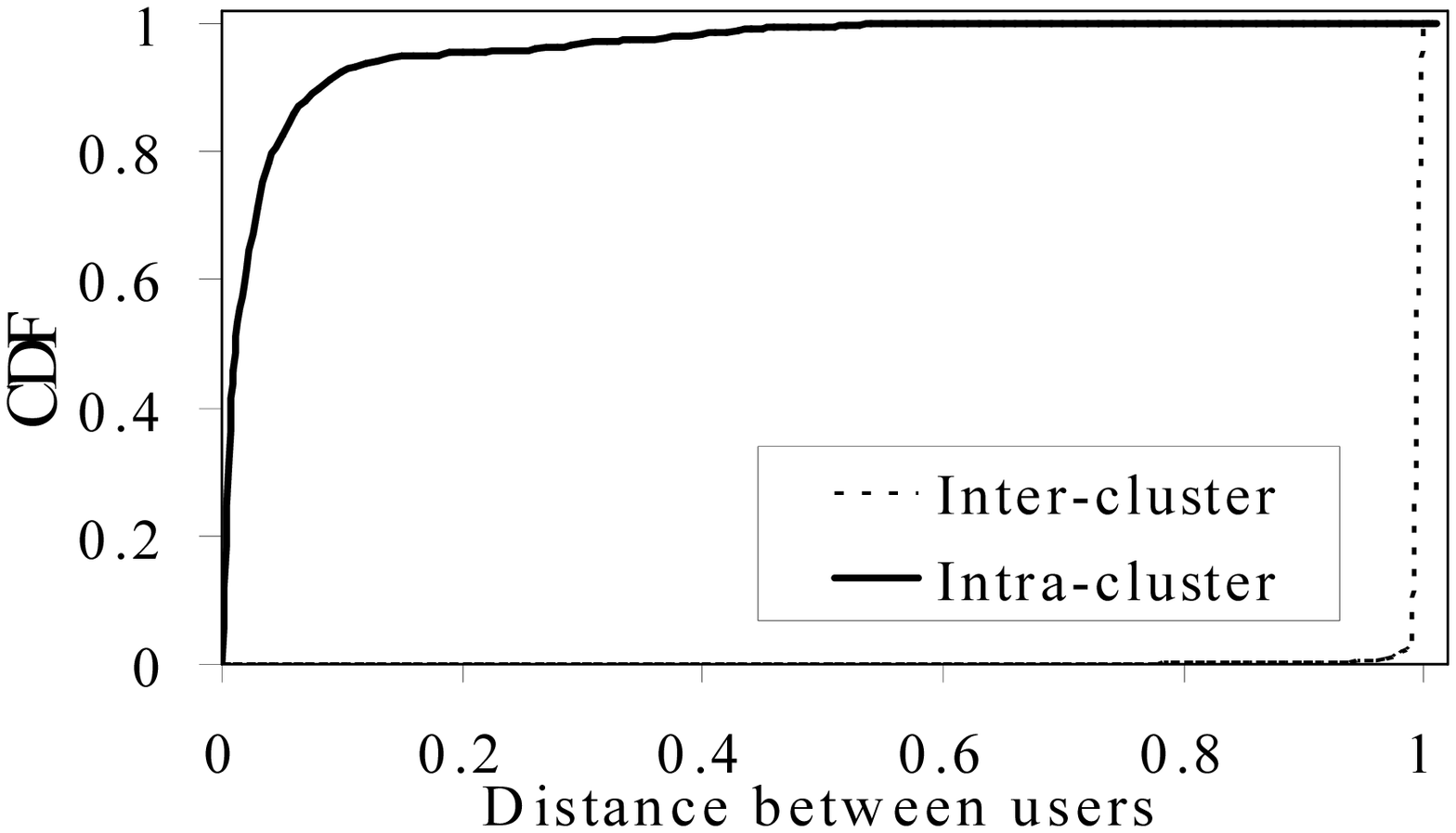} 

\footnotesize{(b) Dartmouth, distance between centroid of the first behavioral mode.}

\end{minipage}
\hfill

\caption{Cumulative distribution function of distances for inter-cluster and intra-cluster user pairs (other distance metrics).}

\label{inter-intra-dist-other}
\end{figure}

We establish a meaningful partition of both user populations in Fig. \ref{inter-intra-dist-feature} with 
the eigen-behavior distance.
However, we have to note that the other summaries presented in section \ref{summary-proposals} could also be used to 
obtain distance metrics. For these single-vector summaries,
such as the average of association vectors ($X_{onavg}$, Eq. (\ref{onavg})) or centroid of the first cluster ($X_{centroid1}$, Eq. (\ref{centroid1})),
we define distance metrics between users by simply calculating the Manhattan distance (Eq. (\ref{man-distance})) between
the corresponding summary vectors.
With these distance metrics, we could also arrive at meaningful partitions of user populations
and hence those are valid metrics, too.
We show two such examples in Fig. \ref{inter-intra-dist-other} - the general observation is that
while $X_{onavg}$ leads to less well-separated clusters than the eigen-behavior distance,
$X_{centroid1}$ leads to even better results.

We need to further compare these different partitions of the user population to understand their
properties.
We choose to use the Jaccard index \cite{jaccard} to compare the similarity between different
partitions of the same population. The Jaccard index between two partitions on the
same population is defined as
\begin{equation}
J(P, Q) = r/(r+u+v),
\end{equation}
where $r$ is the number of user pairs who are partitioned in the same cluster in both partition $P$ and $Q$ (i.e., where
the two partitions agree on the classification). $u$ (or $v$) is the number of pairs who are in the same cluster in $P$ (or $Q$)
but in different clusters in $Q$ (or $P$) (i.e., where the two partitions disagree). We choose the Jaccard index among
many other indices for partition similarity due to its low variance on partitions with the same transfer distance \cite{jaccard}.
For both traces, we list the Jaccard indices between user partitions from various distance metrics (Average and the Centroid
of first behavioral mode) and the partition from the eigen-behavior distance in Table \ref{jaccard_compare}. The Jaccard indices are mostly in the range of $0.7$ to $0.8$, indicating the partitions are in fact similar. The better separation between intra and inter cluster distance distributions with the $X_{centroid1}$ metric is partly because the distances are calculated based on a subset of association vectors with a coherent trend, discarding other vectors. Nonetheless, different distance metric has its own emphasis. While we argue in section \ref{interpretation} with an example that eigen-behavior distance is useful for classifying users with multiple frequently visited locations with similar preferences, this is not always the only goal. Depending on the application, sometimes one may want to consider only the first behavioral mode and ignore the others.

\begin{table}
\caption{Jaccard indices between user partitions based on eigen-behavior distances and various distance metrics.}
\label{jaccard_compare}
\begin{center}
\begin{tabular}{|c||c|c|c|}
\hline
Distance & \mr{Average} & Centroid w/ & Centroid w/  \\
metric & & threshold = 0.5 & threshold = 0.9 \\
\hline
USC & 0.757 & 0.741 & 0.696 \\
\hline
Dartmouth & 0.801 & 0.706 & 0.710 \\
\hline

\end{tabular}

\end{center}
\end{table}

Instead of applying SVD, one may propose to use the centroids for all behavioral modes of a
user as a summary. However, the behavioral mode for each user is dependent
on the clustering threshold, and it is not simple to choose one that works well for many users,
considering the diversity. On the other hand, SVD does not require parameter tuning, and is
optimal in the sense of capturing remaining power in the association matrix, so we choose it over
the multiple centroids method, if all behavioral modes of a user should be considered.

\subsection{Various Data Representations} \label{alternative-representation}

In this section we discuss about alternative representations of user behavior and compare the
findings with the {\it normalized association vector} we choose in the paper.

In section \ref{representation} we propose to use normalized association vector in order to mitigate the differences
of user activeness across users and across time slots for a given user. This is effective if the {\it preference}
of the user for each time slot is the focus of study. For example, if a user visits exclusively location $A$ or $B$ on different
days, for similar number of days, one way to understand the user behavior is that locations $A$ and $B$ are of the
same importance for the given user, since he visits these two places exclusively with similar frequency. However,
if the user stays at location $A$ whenever he visits for much longer duration than $B$, the normalized vector would
not reveal such information. On the other hand, if absolute association time is used in the vectors, the large time spent
at location $A$ will hide his visits to $B$ when SVD is applied to extract eigen-behaviors from the user (i.e., vectors
with large association time to $A$ dominate the power of the matrix), albeit the user pays a lot of visits to $B$.

Both representations may be of interest to some applications. So instead of arguing the importance of one
over the other, we try to understand its impact on how the users are clustered. Using the USC trace as an example, we 
compare the results of user partitions using absolute association time vectors and normalized association vectors.
We observe that the most active users (i.e. The first quarter in terms of online time)
are almost classified the same regardless which
representation is used, with Jaccard index $0.9652$. This is the case since the most active users are almost always
on, and the representation does not make much difference. We see the Jaccard indices drop to $0.7910$, $0.7090$,
and $0.6096$ for the second, third, and fourth quarter of users in terms of activeness, respectively, a clear decreasing
trend. The least active users are more sensitive to the choice of representation due to their sporadic usage of WLAN.

Time slot sizes to collect the association vectors is another dimension to experiment with.
In addition to daily vectors, we consider two other schemes: (1) Generate association vectors
for every three-hour time slot. We compare the partitions generated by this fine-grained representation
with the partition generated by the daily representation, and get the Jaccard indices of
$0.787$ and $0.778$ for USC and Dartmouth, respectively. This indicates that a finer time scale
does not change the user classification much, and one day interval would be sufficient
to capture important trends in user behavior. We also try (2) generate association vectors only
during the time frame between 8AM to 4PM, the busy part of a day, and compare the subsequent
partition with the partition generated by the daily representation in which the whole day is included. With this representation,
the two traces give very different result - the Jaccard indices are $0.752$ and $0.033$, respectively.
Hence it is not always sufficient to use only the behavior trends during working hours to classify
users.

The choice of location granularity in the representation is important to understand the results.
We have also try to use access points as locations for Dartmouth trace as the information is
available. For most of the studies, the observations are similar to what we present in the paper, although one can expect
to see more distinct behavior groups from the population if finer location granularity is used.
However, it is not easy to interpret these groups meaningfully unless we have the information
about detailed AP locations within the buildings and the significance of its covered area in social
context. On the other hand, it is also possible that
a group of buildings bear a higher-level meaning in social context (e.g., Several close-by dorms form
a "residential area", or close-by buildings shared by the students from the same department), and
it is also related to understand user visiting preferences from a higher-level behavioral context (e.g.,
home, at work, at class, etc.). We leave this as future work.

\end{document}